\documentclass{aa}  

\usepackage[normalem]{ulem}

\usepackage{graphicx}
\usepackage{caption}
\usepackage{txfonts}
\usepackage{color}
\usepackage{subfigure}
\usepackage[dvipsnames]{xcolor}
\usepackage{longtable} % for 'longtable' environment
\usepackage{pdflscape}
\usepackage{afterpage}
\usepackage{capt-of}% or use the larger `caption` package
\usepackage{soul}
\begin{document}

   \title{The stellar host in star-forming low-mass galaxies: Evidence for two classes}

   \author{A. Lumbreras-Calle
          \inst{1,2}\fnmsep\thanks{\email{alcalle@iac.es}}
          \and
          J. Méndez-Abreu\inst{1,2}
          \and
          C. Muñoz-Tuñón\inst{1,2}
          }

   \institute{Instituto de Astrofísica de Canarias, Calle Vía Láctea s/n, E-38205 La Laguna, Tenerife, Spain\\
         \and
             Universidad de La Laguna, Departamento de Astrofísica, Avenida Astrofísico Francisco Sánchez, s/n, E-38206 La Laguna, Tenerife, Spain\\}

   \date{Received}

 \abstract{The morphological evolution of star-forming galaxies provides important clues to understand their physical properties, as well as the triggering and quenching mechanisms of star formation.}{We analyze the morphology of galaxies hosting star-forming events at low redshift ($z<0.36$). We aim at connecting morphology and star-formation properties of low-mass galaxies (median stellar mass $\sim$ 10$^{8.5}$ M$_{\odot}$) beyond the local Universe.}{We use a sample of medium-band selected star-forming galaxies from the GOODS-North field. H$\alpha$ images for the sample are created combining both spectral energy distribution fits and HST data. Using them, we mask the star forming regions to obtain an unbiased two-dimensional model of the light distribution of the host galaxies. For this purpose we use \texttt{PHI}, a new Bayesian photometric decomposition code. We apply it independently to 7 HST bands, from the ultraviolet to the near-infrared, assuming a Sérsic surface brightness model.}{Star-forming galaxy hosts show low Sérsic index (with median $n$ $\sim$ 0.9), as well as small sizes (median R$_e$ $\sim$ 1.6 kpc), and negligible change of the parameters with wavelength (except for the axis ratio, which grows with wavelength in 46\% of the sample). Using a clustering algorithm, we find two different classes of star-forming galaxies: A more compact, redder, and high-$n$ (class A) and a more extended, bluer and lower-$n$ one (class B). This separation holds across all seven bands analyzed. In addition, we find evidence that the first class is more spheroidal-like (according to the distribution of observed axis ratios). We compute the color gradients of the host galaxies finding that 48\% of the objects where the analysis could be performed show negative gradients, and only in 5\% they are positive.}{The host component of low-mass star-forming galaxies at $z<0.36$ separates into two different classes, similar to what has been found for their higher mass counterparts. The results are consistent with an evolution from class B to class A. Several mechanisms from the literature, like minor and major mergers, and violent disk instability, can explain the physical process behind the likely transition between the classes.}
% 5 {} token are mandatory
 % \abstract


   \maketitle
%
%________________________________________________________________
\section{Introduction}
\label{sec:intro}

During the last decades, numerous studies have explored the different morphological, structural, and colour properties of star-forming galaxies (SFGs). The relation between both the star formation (SF) and structural properties of galaxies, as well as their dependence with galaxy mass and cosmic time are key to understand galaxy evolution. Large surveys show that the bulk of star formation happens in disk galaxies, as defined by their "near-exponential" surface brightness profiles \citep{Brennan17}, with stellar masses below 10$^{10.5}$ M$_{\odot}$, and  stellar mass surface densities lower than 10$^{8.8}$ M$_{\odot}$ kpc$^{-2}$ \citep{Kauffmann03,Ilbert13}.

In the nearby universe, HII galaxies, Blue Compact Dwarfs (BCDs), and dwarf irregular galaxies (dIrr) dominate the low-mass star-forming galaxy population. In order to understand the possible relation between the structure and star formation processes detailed photometric studies of SFGs revealed the necessity of either masking the star-forming regions \citep{Cairos02, Cairos03, Caon05, Amorin09} or analysing only the outermost parts of the galaxies \citep{Janowiecki14}. These works unveiled old, disk-like host galaxies underneath the starburst events that dominate the optical emission of these systems.

At higher redshifts ($z>0.5$), the morphological analysis of SFGs faces problems such as Malmquist bias, cosmological dimming, and lack of the needed spatial resolution \citep[][and references therein]{Paulino-Afonso17}. This implies that galaxies analyzed in these studies are typically more massive (M$_\star$> $10^{9.5}$ M$_\odot$) than those at lower redshifts. Still, most of the previous work at these redshifts show that star formation tends to happen in disk galaxies with Sersic index $n \sim 1$ \citep[e.g.][]{Wuyts11b}. Even compact, starbursting systems, such as Green Pea galaxies have been shown to host an underlying disk-like stellar structure similarly as their local counterparts \citep{Amorin12}. As redshift increases, the typical specific star-forming rate (sSFR) increases as well due to the SFR being higher for galaxies with the same stellar mass \citep{Oliver10,Whitaker14,Tasca15}. Galaxy morphology also changes at higher redshift, with galaxies showing large and massive star-forming knots \citep{Abraham96,vandenBergh96, Elmegreen04, Elmegreen04b, Elmegreen07}. Similarly as in low-redshift systems, an accurate description of the structural properties of the host galaxy can only be achieved once a separated analysis of the clumps (the current star formation) from that of the host galaxy (previously formed stars) is carried out. It has been also shown that star-forming clumps closer to the center of a galaxy are generally older, redder, denser (Guo et al. 2012), and more massive \citep{Hinojosa-Goni16, Cava18} than those at larger galactocentric distances. This supports the idea that these massive clumps are created through violent disk instabilities  \citep{Bournaud07}. Their coalescence and infall to the galaxy center eventually gives rise to central concentrations of stars thus reshaping the morphology of the galaxy \citep{Ceverino10}.

The structural and morphological properties of galaxies are also linked to the triggering and quenching of star formation. Indeed, several observational works have pointed out the existence of a transitional population of galaxies \citep{Dellenbusch08,Meyer14,Lian15,Koleva13,Pandya17,Wang17,Wang18,Kelvin14,Pawlik18,Maltby18}. They show intermediate properties (in size, Sérsic index, star formation rate) between passive and normal SFGs, thus hinting to a possible evolutionary path where these properties are connected. In low-mass galaxies it has been extensively determined, both observationally and in simulations, that the star formation history (SFH) of low-mass galaxies is more bursty than in higher mass galaxies \citep{DiMatteo08,Bauer13,Sparre17,Faucher-guiguere18}. The triggering mechanisms can be due to accretion of pristine gas \citep{SanchezAlmeida13,SanchezAlmeida15}, the compression of hot gas around the galaxies due to interactions with the intergalactic medium \citep{Wright19}, or encounters with other galaxies \citep{Stierwalt15}. The quenching processes at these masses is thought to be dominated by the environment where they reside \citep{Fillingham18}. At high galaxy masses, more complex quenching mechanisms have been proposed, which may still play a role in the lower mass range. The {\it compaction} scenario suggests that galaxies grow their central stellar densities (through minor/major mergers, or violent disk instability) compacting the galaxy, and stopping star formation as a result \citep{Dekel14}. Along this line, another possibility is the {\it morphological quenching} theory \citep{Martig09}, where the growth of central stellar mass density inhibits further star formation by stabilizing the gas in the disk against gravitational collapse. The analysis of the structure of SFGs, and their diverse properties can help constrain the possible evolutionary pathways between them and the physical reasons driving these changes.

In this work, we present an accurate modeling of the surface brightness distribution and structural properties of the host component of low mass emission line galaxies (ELGs) at low redshift ($z < 0.36$). We aim at filling the gap between large surveys (at masses higher than $10^9$ M$_{\odot}$) and local dwarf galaxies analysis, with particular emphasis in the presence of transitional types (i.e., galaxies transforming from star forming to passive) at this masses and redshifts. The sample, obtained in \cite{Lumbreras-Calle18} (hereafter referred to as Paper I) includes low-mass star-forming system identified thanks to the use of data from the Survey for High-z Absorption Red and Dead Sources \citep[SHARDS,][]{Perez-Gonzalez13}, a deep, medium-band multi-wavelenght survey. In order to robustly analyse the morphological properties of the galaxies, we use the high-resolution, deep HST images from the CANDELS (Koekemoer et al. 2009) and 3D-HST (Skelton et al. 2014) surveys. 

The paper is structured as follows. Section 2 describes the sample selection and the code used in the photometric decomposition. Section 3, describes the process of masking the star-forming regions and performing the actual photometric fits. In Section 4 we analyse the results of the photometric fits in the context of the galaxy properties. Section 5 discusses the results and Section 6 presents our conclusions. 

Throughout this paper we consider standard $\Lambda$CDM cosmology, with $\Omega_{\Lambda}$=0.73, $\Omega_{M}$=0.27 and $H_{0}=71$ km s$^{-1}$ Mpc$^{-1}$. All uncertainties reported refer to the limits of the central 68\% of the probability distribution.

\section{Sample selection and photometric decomposition}
\label{sec:data}
%\textcolor{black}{Database, sample selection and \texttt{PHI} fitting code}
The initial sample of young SFGs analyzed in this paper is extracted from Paper I. They identified 160 emission line galaxies (ELGs) at $z < 0.36$ within SHARDS \citep{Perez-Gonzalez13}. The SHARDS survey consists of deep imaging of the GOODS-North field using 25 contiguous medium band filters, and it was performed using the OSIRIS \citep[Optical System for Imaging and low-Intermediate-Resolution Integrated Spectroscopy,][]{Cepa00} instrument, at the 10.4 m Gran Telescopio de Canarias (GTC) at the Observatorio del Roque de los Muchachos, in La Palma. 

We refer the readers to Paper I for a full description of the sample selection and characterization. Briefly, they developed an algorithm to simultaneously detect the H$\alpha$ and [OIII] emission lines, as star formation tracers, in the well sampled ($R \sim 50$) SED provided by SHARDS. They removed contamination by active galactic nucleii from the sample, ensuring that the emission lines come from star-forming events. In addition, models with two single stellar populations (SSP), one young and one old, were used to fit the SED of the galaxies and to obtain their stellar population properties. Thanks to the depth and quasi-spectroscopic resolution of SHARDS they were able to detect SFGs with equivalent widths (EWs) as low as 12 \text{\AA} (median values of $\sim$35 \text{\AA} in [OIII] 5007 and $\sim$56 \text{\AA} in H$\alpha$) and stellar masses as low as 10$^{7} $M$_{\sun}$ (with median value of 10$^{8.5}$ M$_{\sun}$).

The SHARDS survey covers the Hubble Space Telescope (HST) legacy field GOODS-North. Thus, multi-wavelength high-spatial resolution imaging is available for all 160 ELGs in the sample. In particular, in this paper we use images in the $F606W$ and $F850LP$ filters from the Advanced Camera for Surveys (ACS) obtained from the CANDELS survey \citep{Grogin11}, the $F435W$ and $F775W$ filters from ACS, and the $F125W$, $F140W$, and $F160W$ filters from the Wide Fide Camera 3 (WFC3) retrieved from the 3DHST survey \citep{Brammer12,Skelton14}. This is the dataset we use to perform our photometric decomposition in the next subsection.

\subsection{\texttt{PHI} photometric decomposition code}
\label{sec:PHI}

Since the development of two-dimensional (2D) photometric decomposition programs such as GALFIT \citep{Peng02}, BUDDA \citep{deSouza04} or GASP2D \citep{Mendez-Abreu08}, vast amount of work has been performed with them, and the limitations in their application to SFGs have been studied in detail \citep[][among others]{Paulino-Afonso17,Amorin09}.
However, some problems related with the algorithms still exist, such as local minima trapping of the solution and inaccurate derivation of the final errors. Therefore, a new approach is needed to provide an accurate description of the stellar host in ELGs at intermediate redshift. In order to overcome some of these difficulties, we used the  \texttt{PHI} code described in \cite{Argyle18}. \texttt{PHI} is based on an adaptive Markov Chain Monte Carlo (MCMC) algorithm that efficiently explores the parameter space providing robust results and statistically meaningful errors. 

A complete description of the code can be found in \cite{Argyle18}. Here we summarize its key aspects. As most 2D fitting codes, \texttt{PHI} uses both the flux and error images of a galaxy, as well as the point-spread-function (PSF). The code starts exploring the parameter space and in each iteration it produces a model image for the galaxy. Then, this model is convolved with the PSF and (given its Bayesian framework) the likelihood and posterior probability are calculated for the model and the data. After that, the MCMC engine starts varying the parameters, repeating the model creation and likelihood and posterior probability computation until the code converges or a maximum number of iterations is reached.

The exploration of the parameter space is performed in three different levels:
\begin{enumerate}
\item \textbf{Blocked Adaptive Metropolis.} In this stage, \texttt{PHI} works by fixing all parameters but one, and only modifies this parameter in order to estimate its typical scale of variation. This makes more efficient the further exploration of the parameter space.
\item \textbf{Adaptive Metropolis.} The code varies all parameters in each iteration to obtain a good description of the covariance matrix.
\item \textbf{Chain convergence.} Three Markov chains are run simultaneously, using the last covariance matrix from level 2, in order to perform the exploration of the parameter space in the most efficient way. We only consider a fit as successful when at least two of these chains converge and give compatible results.
\end{enumerate}

In this paper we use a single S\'ersic function \citep{Sersicdg} to describe the surface brightness distribution of the host galaxy:

\begin{equation}
I(R)=I_e exp \left\{ -b_n \left[ \left(\frac{R}{R_e}\right)^{1/n}-1 \right] \right\}
\end{equation}

where $I_e$ is the intensity at the effective radius $R_e$, enclosing half of the total light of the model. $n$ is the Sérsic index, describing the concentration of the light profile, and $b_n$ is a parameter determined by $n$. This model is implemented in concentric elliptical isophotes with position angle $\theta$ and ellipticity $\epsilon=1-q$, where $q=b/a$ is the ratio between the semi-minor and semi-major axis of the ellipse. It is worth noting that we use uniform priors in all parameters involved in the fit. In the case of $I_e$ and $R_e$ the fit is performed in logarithmic units. The Sérsic index $n$ is allowed to vary between 0.4 and 8, and $q$ between 0.2 and 1.
 
\section{Masking and fitting process}
\label{seccion_mask_fit}

To robustly describe the surface brightness distribution of the host galaxies in our sample of ELGs, we first mask out the star-forming regions in the HST images. Not masking them would add spurious flux, possibly biasing the result of the fit \citep[see][and references therein]{Amorin09}. The full process of masking and fitting is summarized in Fig. \ref{fig:flowchart} and described in this Section. We create the masks running the \texttt{SExtractor} code \citep{Bertin96} with different threshold levels on H$\alpha$ images created using HST broadband data. For those galaxies where no clear H$\alpha$ was detected, we create the masks based on the near-UV HST data (F435W filter). Then, we perform the photometric decompositions, remove the poor fits, and choose the appropriate masks (see Fig. \ref{fig:flowchart}).

\subsection{Creation of masks for the star-forming regions}
Our first step is to create {\it pure} H$\alpha$ images using the available broad band HST images. To this aim, we use data from two HST filters, one of them including the H$\alpha$ line emission, and the other probing only the stellar continuum. The process starts by aligning and PSF-matching both images. Then, we subtract the continuum image from the emission line one. In particular, for galaxies at $z$<0.266, the H$\alpha$ line lies in the F775W filter and F850LP was used as continuum. For those with $z$>0.298, H$\alpha$ lies in F850LP and F775W contains the continuum. In between those redshifts, the process is more complex since H$\alpha$ is covered with both filters (see Appendix \ref{appen_imag} for more details). 

As a simple subtraction of both images (with and without H$\alpha$) is not accurate enough, we take into account the continuum variation between the two filters, using the best fit models derived in Paper I. These were obtained by fitting stellar population models to SHARDS data, which provides higher spectral resolution than HST, thus making this process feasible. To include the uncertainties in the SED fitting, we perturb the scaling factor ($\pm$10\%) between filters choosing the value that produces less continuum oversubstraction when generating the H$\alpha$ images. 

Once the {\it pure} H$\alpha$ images are built, the mask creation process consists in detecting sources over a given threshold. We run \texttt{SExtractor} with different thresholds over the sky background (1.1$\sigma$, 1.5$\sigma$, and 2 $\sigma$). The output segmentation maps were directly used as star-forming region masks. Since our ELG detection process (described in Paper I) is based on the SHARDS survey, which is very efficient detecting emission lines, some galaxies show no measurable H$\alpha$ in the HST images. In these cases we use the near-UV HST data (F435W filter) to select the most prominent UV-emitting regions as masks.

\subsection{Quality check and mask selection}

\begin{figure}
   \centering
   \includegraphics[width=0.5\textwidth,keepaspectratio]{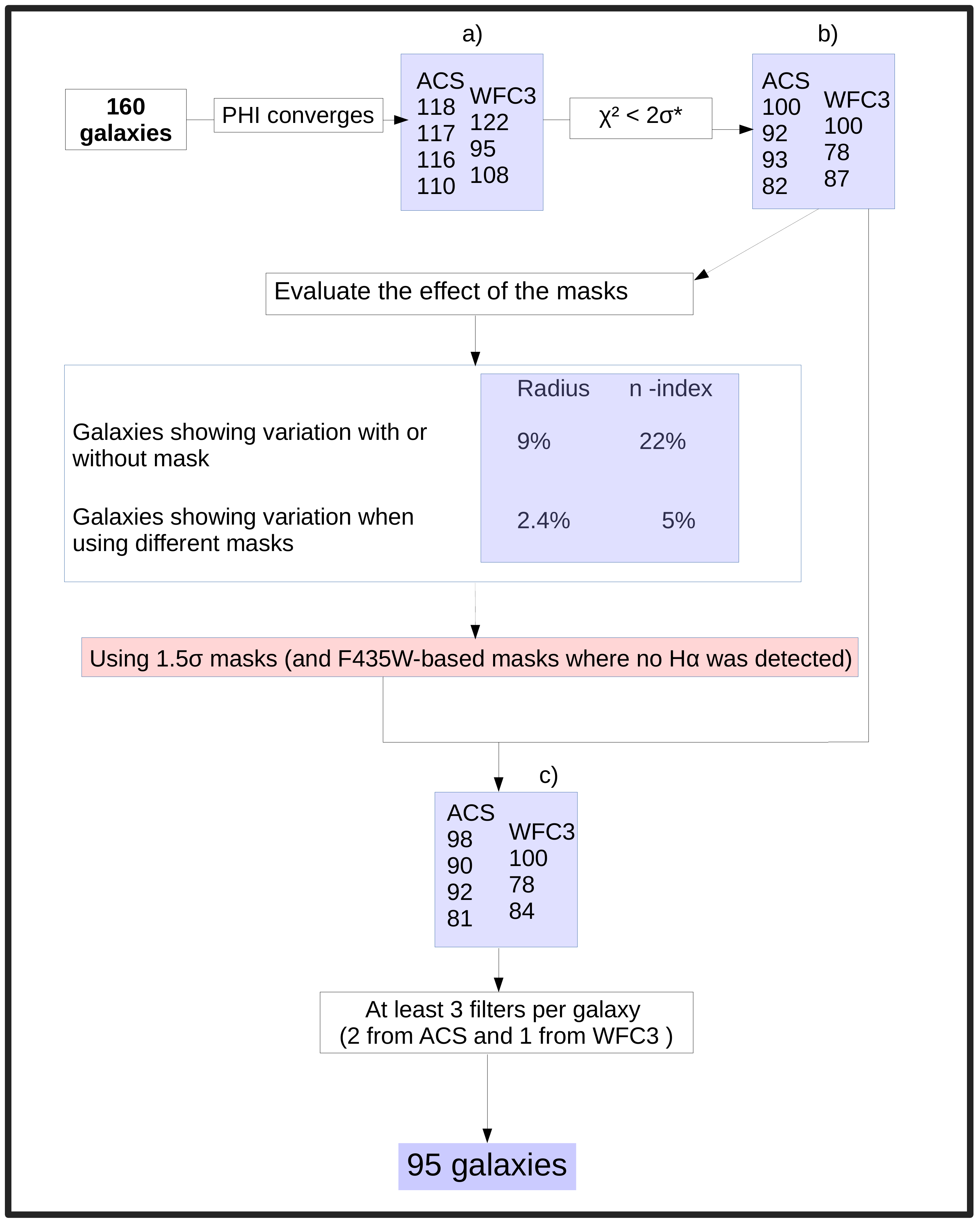}
      \caption{Flowchart summarizing the process of choosing the appropriate fits and masks. The numbers in boxes a) and b) represent the median amount of galaxies in the sample, considering all masks, for each one of the four filters in ACS (from top to bottom, F435W, F606W, F775W, and F850LP) and the three in WFC3 (F125W, F140W, and F160W). The quantities in box c) represent the number of galaxies in each band fulfilling all criteria, after choosing the final mask.}
         \label{fig:flowchart}
   \end{figure}

\begin{figure}
   \centering
   \includegraphics[width=0.5\textwidth,keepaspectratio]{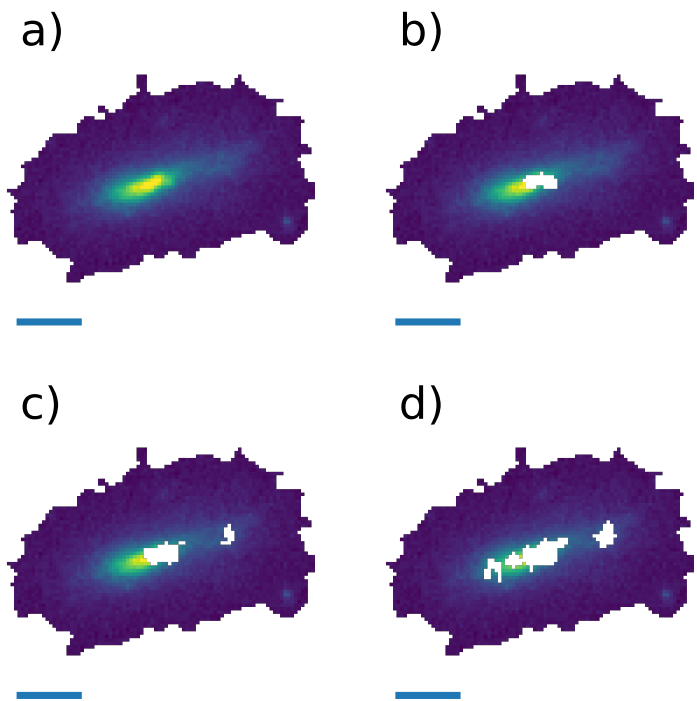}
      \caption{Example of a galaxy from our sample without any mask (panel a) and with the three different H$\alpha$ detection masks over it: 2$\sigma$ (b), 1.5 $\sigma$ (c) and 1.1 $\sigma$ (d). The horizontal line represents a 1 kpc distance. }
         \label{fig:examp_mask}
   \end{figure}

All galaxies in our sample are photometrically analyzed using \texttt{PHI} in the 7 HST filters ($F435W$, $F606W$, $F775W$, $F850LP$, $F125W$, $F140W$, $F160W$). We use 4 different masking modes (no mask, 1.1$\sigma$, 1.5$\sigma$, and 2$\sigma$ over the threshold mask). An example of the different masking regions is shown in Figure \ref{fig:examp_mask}. In total, we carry out 4480 (160 galaxies x 7 bands x 4 masks) photometric decompositions using the \texttt{PHI} code. Most of these fits converge, but a few either failed or showed high $\chi^2$ values, and they were removed from the sample. In order to choose the most accurate one, we evaluate the results obtained for different masks. The whole process is summarized in Fig. \ref{fig:flowchart}, where we also show the amount of galaxies retained in each step for each band, and the effect of the different mask modes in the fitted values of radius and Sérsic index.

 \begin{figure*}
   \centering

   \includegraphics[width=0.95\textwidth,keepaspectratio]{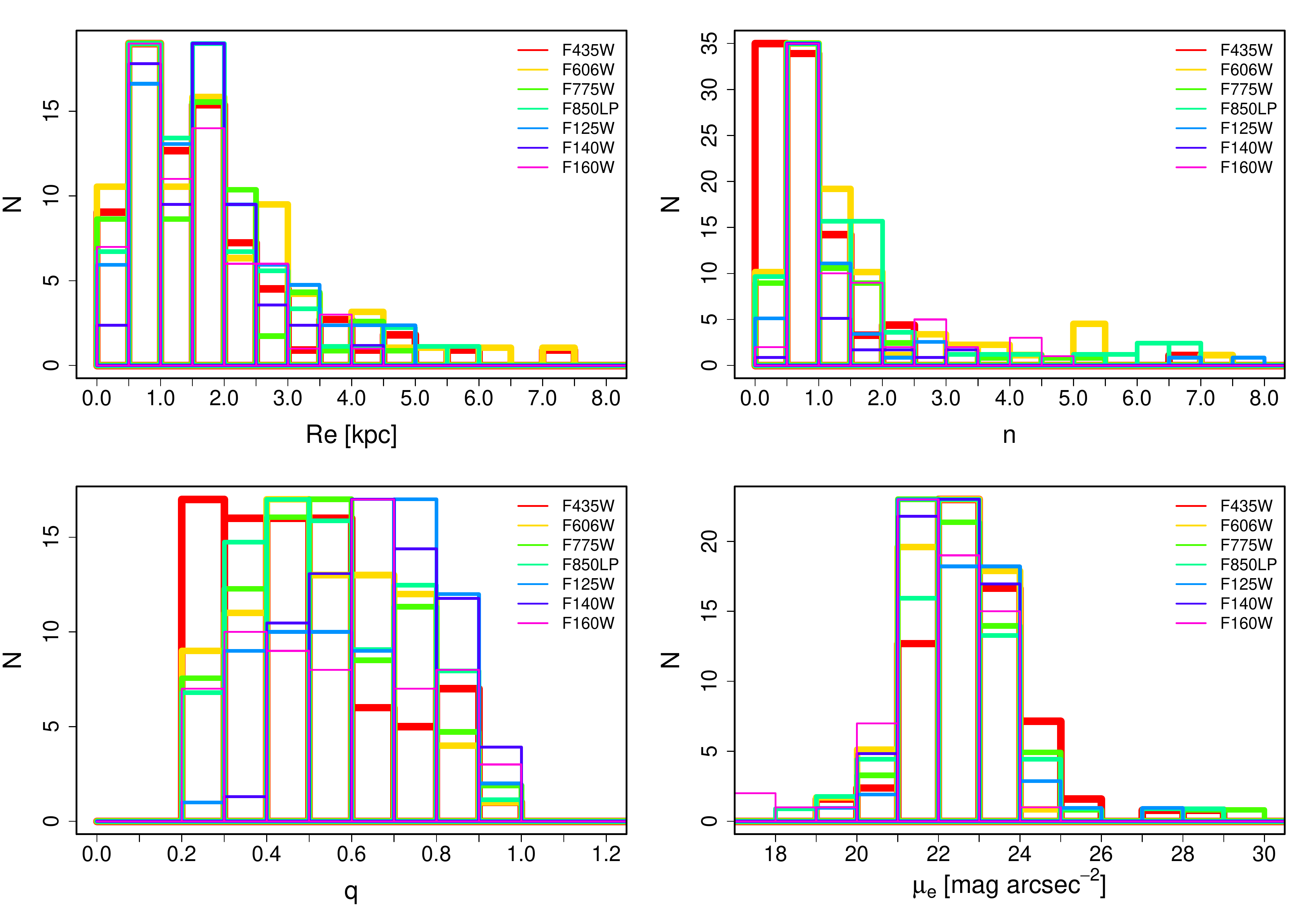}
      \caption{Distribution of the main photometric parameters derived for our ELG sample. The upper left panel presents the values of the effective radius of the best-fit Sérsic model for the galaxies in each filter, in kpc. The upper right panel shows the histogram of the Sérsic index, the lower left panel the axis ratio $q$, and the lower right panel the surface brightness at the effective radius (after applying the cosmological dimming and K corrections). }
         \label{fig:resul}
   \end{figure*}

\subsubsection{Removing poor fits}
We find that for some of the galaxies (15 in median, considering the different bands and masks) the code does not reach a solution. This is either due to galaxies being too small, faint, or due to errors in the photometry in certain bands. Among those runs where the code does provide a solution we find $\sim$ 29 (depending on the filter and mask) where either none or only one of the three chains converges successfully. We also remove those fits since we cannot assure they are reliable, and we are left with the amount of galaxies listed in box a) of Figure\ref{fig:flowchart} for each filter. 

In addition, we evaluate the goodness of the model by analyzing its $\chi^2$.  The offset between filters was accounted for by normalizing the values in order to apply the same criterion in all bands. We measure the distribution of $\chi^2$ in each band, defining $\sigma$* as the distance between the 16 and 50 percentiles. After visually inspecting the residual images, we place the threshold at 2$\sigma$* over the median $\chi^2$. The fits presenting higher values are removed from the sample, and the amount of galaxies left in the sample is presented in box b) of Fig \ref{fig:flowchart}.

\subsubsection{Mask selection}

We then analyze the fitted parameters of the galaxies as a function of the mask used to cover the star-forming regions. We find a significant difference between the photometric parameters when the galaxies are fitted with or without mask (independently of which one is used). For instance, we find differences greater than 20\% in 22.5\% and 9.5\% of the galaxies when considering the S\'ersic index $n$ and effective radius $R_e$, respectively. However, these values drop to 5.1\% and 2.4\% when comparing the best fitted values among different masks. This means that the presence or absence of a mask has a noticeable influence in the fit for some of the galaxies, and therefore it is necessary to use them to derive precise morphological properties of the host galaxy. The precise shape and size of the mask has, however, a small impact on the fitted parameters. Therefore, as a compromise, we choose to use our intermediate H$\alpha$ mask (with detection threshold of 1.5$\sigma$ over the sky noise). With this choice, we avoid both large masks that leave very few galaxy pixels to fit and small masks that leave many pixels contaminated by strong star-forming regions possibly biasing the fit.

Using the 1.5$\sigma$ masks, we are left with the amount of galaxies per filter shown in box c) of Fig. \ref{fig:flowchart}. They fulfill the three criteria previously mentioned: \texttt{PHI} finds a solution, two or three chains converge, and the normalized $\chi^2$ of the model is smaller than the threshold. Considering this, 143 galaxies are successfully fitted at least in one band. To select the final sample of galaxies over which we will perform the rest of the analysis, we only keep those that present good fits in at least three bands: two or more filters from ACS, and one or more from WFC3. This is done to ensure that an analysis of the galaxy colors is feasible without extrapolating values to distant wavelength ranges. Applying this criteria, 95 galaxies constitute the final sample studied in this paper.
 \begin{figure}
   \centering
   \subfigure{\includegraphics[width=0.45\textwidth,keepaspectratio]{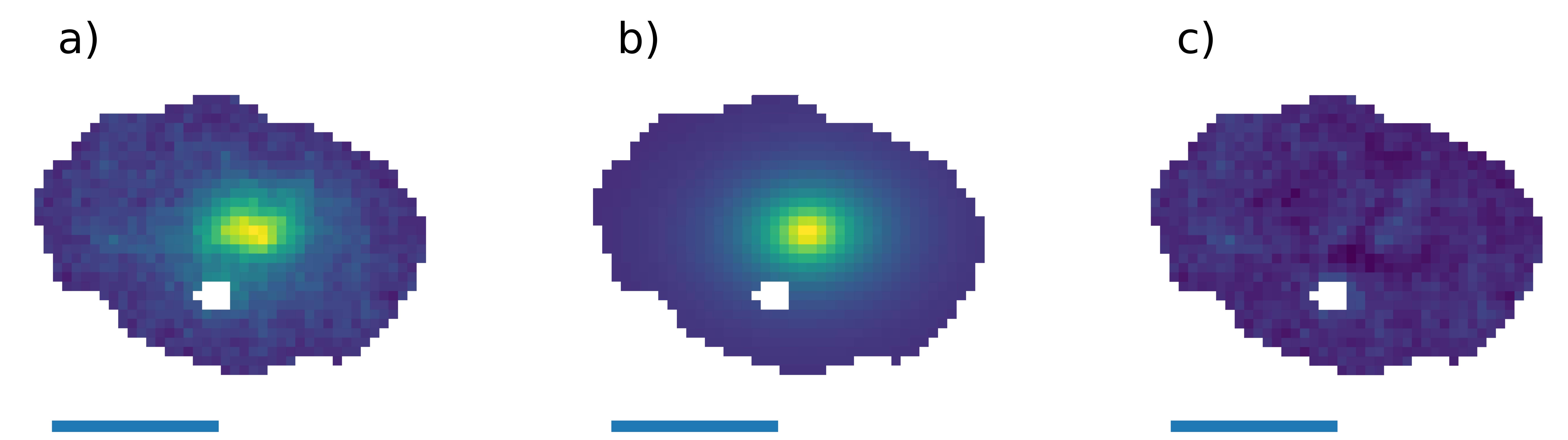}}
   \subfigure{\includegraphics[width=0.2\textwidth,keepaspectratio]{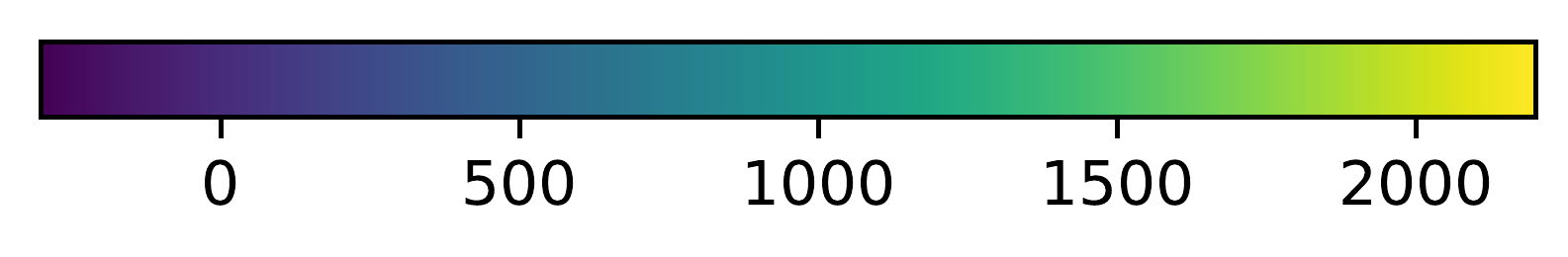}}
   \subfigure{\includegraphics[width=0.45\textwidth,keepaspectratio]{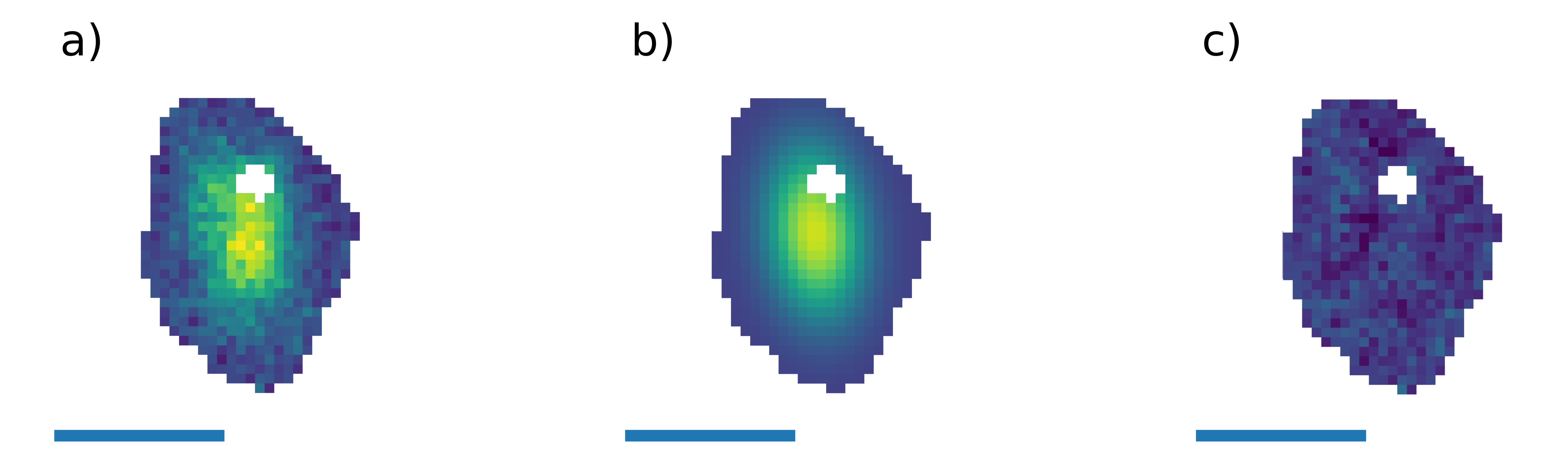}}
      \caption{Example of two galaxies from our sample, with the original image and the H$\alpha$ detection mask overlaid on it (left column), the Sérsic model (central column) and the residuals from the fit (right column). All images are presented with the same color scale, shown in the central color bar (in counts). }
         \label{fig:examp_model}
   \end{figure}

\section{Results}

In this Section we summarize the results of applying our photometric decomposition to the ELGs. In Table \ref{tab:phres_tab} we give the best-fitting values (with uncertainties) for all the parameters in all galaxies and bands, as well as the $\chi^2$ values. The uncertainties in the photometric parameters for each galaxy have been derived using their marginalized posterior distributions. We find that more than 70\% of galaxies present uncertainties lower than 25\% on every parameter in all bands, except for the Sérsic index and Sérsic total intensity in the WFC3 filters. In fact, considering only ACS bands, more than 70\% of galaxies present uncertainties lower than 10\% in all parameters, but the Sérsic index and Sérsic effective intensity. The final sample fitted is composed of 95 galaxies out of the original 160 detected ELGs (60\%). 

\subsection{Photometric properties of the host galaxies}

Figure \ref{fig:resul} presents the distribution of the photometric parameters: effective radius $R_e$, Sérsic index $n$, axis ratio $q$, and effective surface brightness $\mu_e$ for the 95 galaxies analyzed. Different colors represent the results in the different HST bands. 

We find that the median effective radii of the host galaxies are lower than 1.7 kpc in all bands. These values are low compared to some previous studies \citep[e.g.][]{Wuyts11b}, but in agreement with the expectations at the mass range where our sample resides \citep{Lange15,Shibuya15}. Closely inspecting the distribution, we detect a characteristic double-peaked distribution in all bands with maxima at $\sim$0.8 kpc and $\sim$1.9 kpc. We investigate this further in Sect.~\ref{sec:twoclass}.

Regarding the Sérsic index, more than 75\% of the galaxies present values lower than 2 in all bands (and more than 50\% equal or lower than 1), with very few ($\sim$4\%) of them between $3<n<5$. These values of Sérsic index are consistent with the results of other emission-line-selected galaxy samples \citep[e.g.][]{Paulino-Afonso17}. Some specific bands present wider distributions (such as F606W and F850LP), but median values remain very similar, spanning from 0.83 to 1.01 except for the bluest filter (F435W), where the Sérsic index is lower than for the other bands (0.59). 

The axis ratio $q$ of the host galaxies shows a wide range of values (from the 0.2 to 1, the minimum and maximum possible values) in all bands, but with different characteristics in the different filters. The bluest filter (F435W) presents a lower median value (0.46) and flatter distribution than the rest of the bands.

The surface brightness at the effective radius $\mu_e$ presents a similar distribution in all bands, slightly skewed towards brighter values in the WFC3 values. It is important to note that the values in this histogram have been corrected by cosmological dimming and K-correction (following the procedure described in Section \ref{colgal}).

\begin{table*}[t]
\renewcommand{\arraystretch}{1.5}
  \centering
\caption{Photometric parameters obtained from \texttt{PHI} fitting. The full table is available online, only the first rows are shown here as guidance. }
\begin{tabular}{llllllll}
\hline\hline
ID & $ Filter $& $\chi^{2} $& $ \mu _e$& $R_e$ & $n$& $q$& $ PA$ \\
(1) & (2) & (3) & (4) & (5) & (6) & (7) & (8)  \\
\hline

    & $F435W$ &  0.99 & $23.318\substack{+0.065 \\ -0.072}$& $0.326\substack{+0.011 \\ -0.010}$ & $0.679\substack{+0.149 \\ -0.120}$ & $0.565\substack{+0.026 \\ -0.025}$ & $-28.9\substack{+2.1 \\ -2.0}$ \\
    & $F606W$ & 0.83 & $23.121\substack{+0.102 \\ -0.098}$& $0.427\substack{+0.022 \\ -0.021}$ & $1.44\substack{+0.23 \\ -0.20}$ & $0.504\substack{+0.028 \\ -0.028}$ & $-29.0\substack{+1.9 \\ -2.2}$\\
    &  $F775W$  & 0.73 & $22.568\substack{+0.075 \\ -0.070}$& $0.430\substack{+0.006 \\ -0.006}$ & $1.162\substack{+0.054 \\ -0.050}$ & $0.523\substack{+0.008 \\ -0.008}$ & $-25.92\substack{+0.67 \\ -0.70}$\\
   10001022 &  $F850LP$  & 0.95 & $22.407\substack{+0.075 \\ -0.070}$& $0.426\substack{+0.015 \\ -0.014}$ & $1.14\substack{+0.17 \\ -0.14}$ & $0.476\substack{+0.021 \\ -0.022}$ & $-32.4\substack{+1.5 \\ -1.5}$ \\
    &  $F125W$  & 1.8 & $23.04\substack{+0.18 \\ -0.13}$& $0.453\substack{+0.025 \\ -0.024}$ & $1.41\substack{+0.31 \\ -0.26}$ & $0.428\substack{+0.037 \\ -0.036}$ & $-26.4\substack{+2.5 \\ -2.4}$\\
       &  $F140W$  & 1.1 & $22.25\substack{+0.12 \\ -0.13}$& $0.576\substack{+0.054 \\ -0.045}$ & $0.66\substack{+0.24 \\ -0.14}$ & $0.587\substack{+0.064 \\ -0.055}$ & $-25.8\substack{+5.7 \\ -5.8}$\\
          &  $F160W$  & 0.75 & $22.25\substack{+0.12 \\ -0.13}$& $0.432\substack{+0.022 \\ -0.021}$ & $0.80\substack{+0.24 \\ -0.18}$ & $0.506\substack{+0.049 \\ -0.050}$ & $-26.2\substack{+3.1 \\ -3.6}$\\

  \hline
  
\multicolumn{8}{l}{}
 \end{tabular}
\caption*{ \textbf{Notes.} (1) SHARDS ID; (2) HST filter ID; (3) reduced chi-squared of the best-fit; (4) central surface brightness of the host galaxy model, in AB mag arcsec$^{-2}$ (after considering cosmological dimming and K-correction); (5) effective radius of the host galaxy model, in kpc; (6) Sérsic index of the host galaxy model; (7) axis ratio of the host galaxy model; (8) position angle of the host galaxy model, in degrees. }

\label{tab:phres_tab}

\end{table*}

\subsection{Integrated colors of the host galaxies}
\label{colgal}
\begin{figure}
   \centering
   \includegraphics[width=0.5\textwidth,keepaspectratio]{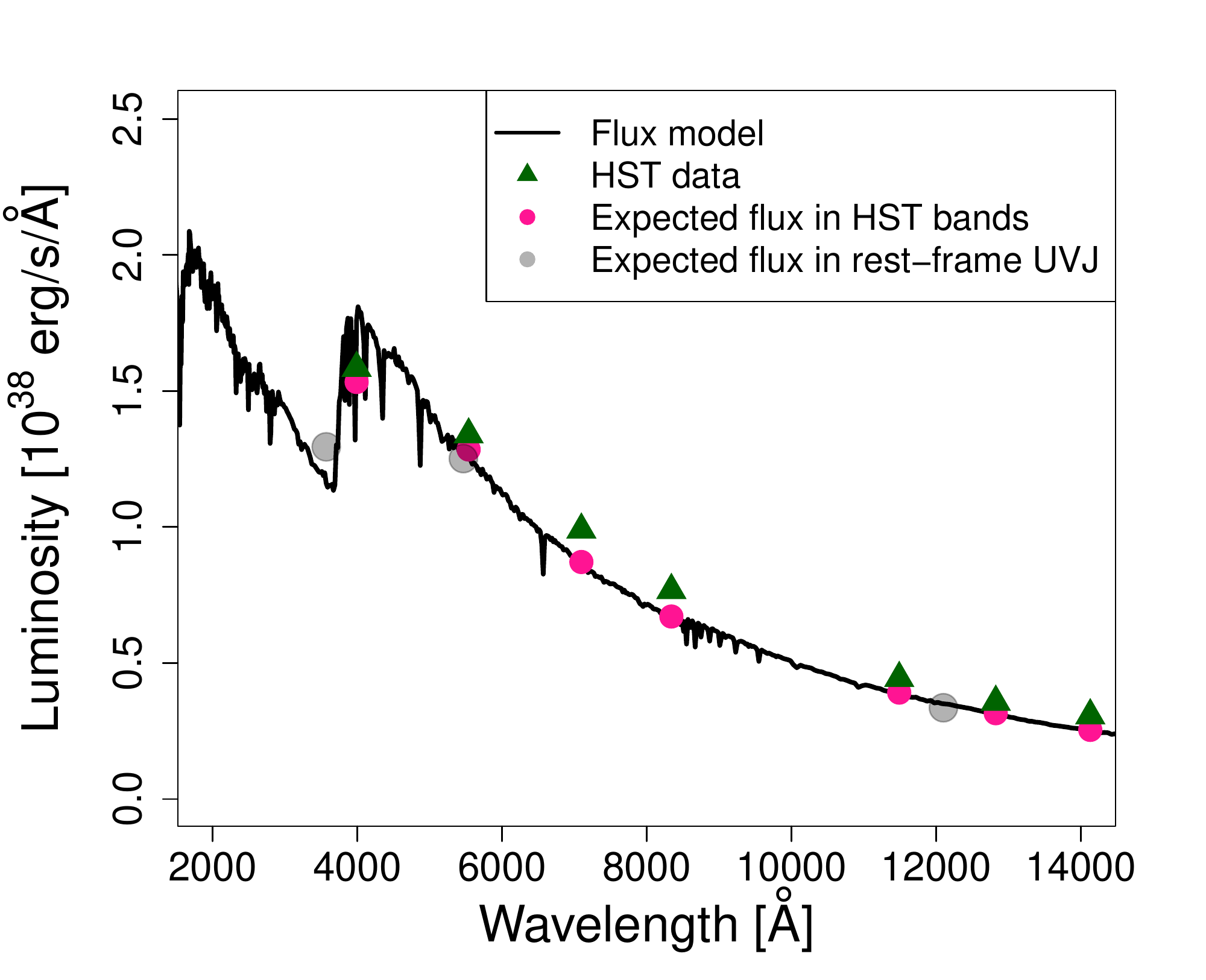}
      \caption{Example of the K-correction process. The composite stellar population model (from Paper I) is shown as a black line. The green triangles represent the integrated flux of the galaxy in each HST band. The pink circles are the expected flux in each HST band, considering only the stellar population model. The grey dots are the expected U, V and J fluxes considering the model. The K-correction is the difference between these fluxes and the expected ones in the nearest HST filters.}
         \label{fig:example_SED}
   \end{figure}

Our multi-wavelength photometric decomposition yields results in seven non rest-frame HST filters, which need to be transformed into standard rest-frame filters to perform color analysis. In order to obtain the U, V, and J rest-frame magnitudes for our models, we apply the K-correction \citep{Hogg02}. To this aim, we take advantage of the SED fits performed on each of the galaxies in Paper I, using the SHARDS data, with higher spectral resolution. We obtain the synthetic magnitudes of the whole galaxies in the U, V, and J rest-frame filters and the HST non-rest frame ones by convolving the population models with the filter transmissions. The difference between the U, V, and J magnitude and the nearest HST band is the K-correction. (see Fig. \ref{fig:example_SED}). We apply it to the integrated magnitude of the Sérsic model of the host galaxy in the appropriate HST bands. 

 \begin{figure}
   \centering
   \includegraphics[width=0.5\textwidth,keepaspectratio]{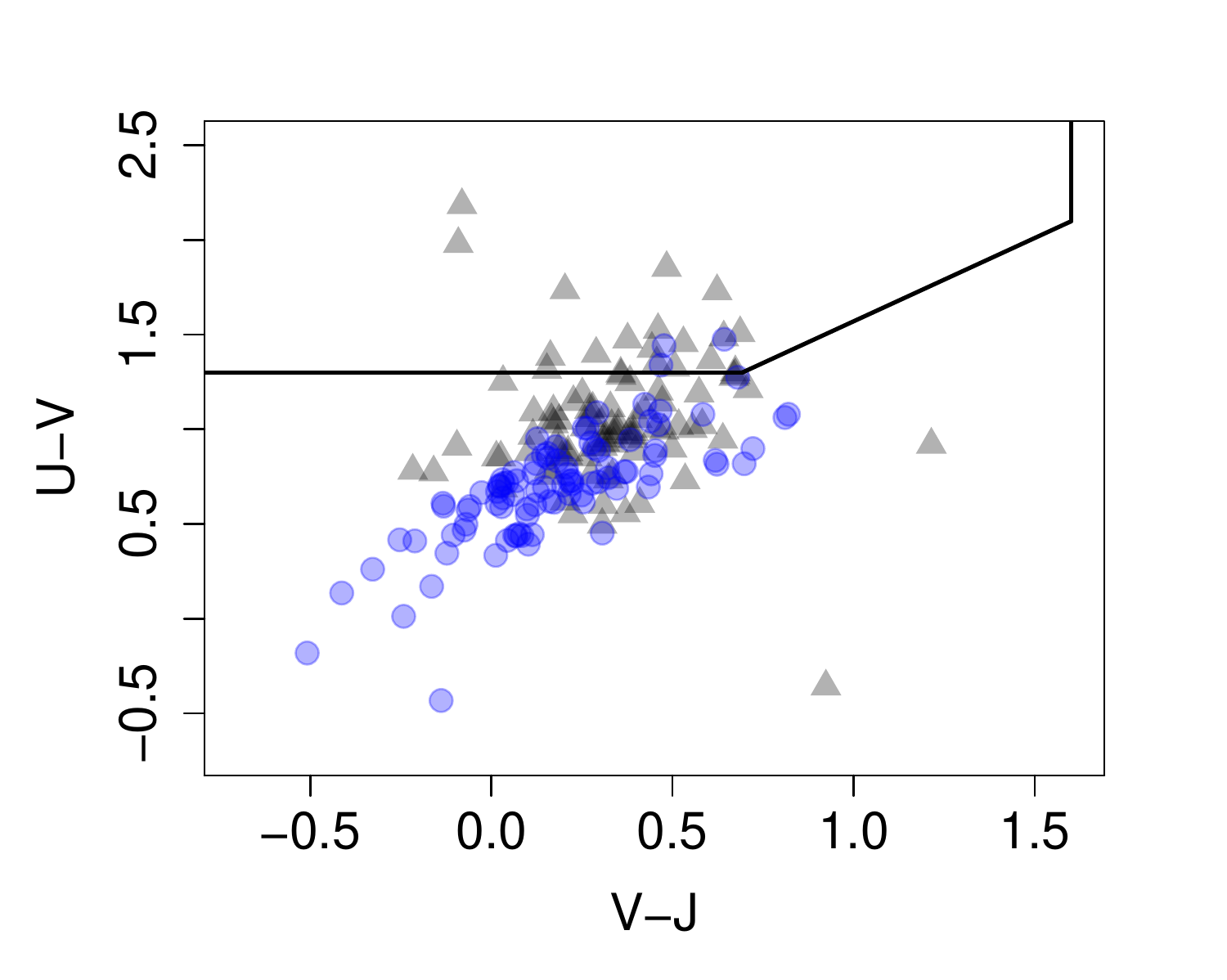}
      \caption{UVJ diagram for the Sérsic models of the host galaxies of the sample (grey triangles). The blue circles represent the integrated colours of the entire galaxies derived from the SED models obtain in Paper I. The black lines represent the separation between passive (top left) and SFGs taken from \citep{Whitaker14}}
         \label{fig:UVJ_diag}
   \end{figure}

Fig. \ref{fig:UVJ_diag} shows the UVJ diagram for the host galaxies (black triangles) and the integrated flux of the whole galaxy, including star-forming regions (blue dots, from Paper I). Despite our sample being exclusively composed of SFGs, we show how some of their hosts actually lie in the passive region of the diagram (top left) once they are free from the star-forming regions.

\begin{table*}[t]
\renewcommand{\arraystretch}{1.5}
  \centering
\caption{Physical properties, synthetic magnitudes, and class to which each galaxy belongs. The full table is available online, only the first rows are shown here as guidance.}
\begin{tabular}{llllllllll}
\hline\hline
ID & Mass & sSFR & U & V & J  & g & r & $\nabla_{g-r}$ & class  \\
(1) & (2) & (3) & (4) & (5) & (6) & (7) & (8) & (9) & (10) \\
\hline

  10001022  & $7.43\substack{+0.17 \\ -0.08}$ & $-9.80\substack{+0.24 \\ -0.56}$ & $26.75\substack{+0.03 \\ -0.04}$& $25.02\substack{+0.04 \\ -0.04}$ & $24.39\substack{+0.09 \\ -0.11}$ & $25.74\substack{+0.03 \\ -0.04}$ & $24.87\substack{+0.04 \\ -0.04}$ & - & A \\
       10000515  & $8.68\substack{+0.01 \\ -0.04}$ & $-9.53\substack{+0.17 \\ -0.83}$ & $24.70\substack{+0.02 \\ -0.02}$& $23.31\substack{+0.01 \\ -0.01}$ & $23.15\substack{+0.04 \\ -0.4}$ & $23.50\substack{+0.03 \\ -0.03}$ & $23.29\substack{+0.01 \\ -0.01}$ & $-0.202\substack{+0.016 \\ -0.016}$ & B \\

  \hline
\multicolumn{9}{l}{} \\
 \end{tabular}

\caption*{ \textbf{Notes.} (1) SHARDS ID; (2) mass of the galaxy, in log(M$_{\odot}$); (3) specific star forming rate of the galaxy, in log(yr$^{-1}$); (4), (5), (6), (7) and (8) AB magnitudes in the U, V, J, g, and r bands, respectively;   (9) g-r color gradient, as defined in Equation \ref{eqn:gradi}; (10) class of the galaxy, (A or B, see text). Data for columns (2) and (3) are taken from Paper I. }

\label{tab:prop_gal}

\end{table*}

\begin{table}
\renewcommand{\arraystretch}{1.5}
  \centering
\caption{Properties of the full sample of galaxies, and of the two separate classes.}\begin{tabular}{llll}
\hline\hline
 & Full sample & Class A & Class B  \\
 \hline
log$_{10}$(M$_{\star}$/M$_\odot$) (1) & $8.45\substack{+0.60\\ -0.78}$ & $8.37\substack{+0.80 \\ -0.82}$  & $8.48\substack{+ 0.51\\ -0.79}$   \\
log$_{10}$(SFR) (2) & $-1.25\substack{+0.65 \\ -0.45}$ & $-1.53\substack{+0.94 \\ -0.39}$ & $-1.20\substack{+0.60 \\ -0.48}$   \\
log$_{10}$(sSFR) (3) & $-9.69 \substack{+ 0.60\\ -0.29}$ & $-9.83\substack{+ 0.65\\ -0.29}$  &  $-9.62\substack{+ 0.55\\ -0.31
}$  \\
$\mu_e$ (4) &  $23.45\substack{+ 1.26\\ -1.05}$ &  $22.74\substack{+ 1.46\\ -1.04}$ &  $23.47\substack{+1.30 \\ -0.87}$ \\
$R_e$ (5) &  $1.49\substack{+1.16 \\ -0.86}$ &  $0.65\substack{+ 0.59\\ -0.28}$ &  $1.82\substack{+1.35 \\ -1.04}$ \\
$n$ (6) &  $0.91\substack{+ 0.90\\ -0.33}$ &  $1.90\substack{+2.55 \\ -0.66}$ &  $0.80\substack{+0.42 \\ -0.29}$ \\
$q$ (7) &  $0.52\substack{+ 0.21\\ -0.17}$ &  $0.53\substack{+ 0.18\\ -0.13}$ &  $0.51\substack{+0.22 \\ -0.19}$ \\
$U-V$ (8) &  $1.01\substack{+0.34 \\ -0.23}$ &  $1.36\substack{+ 0.26\\ -0.67}$ &  $0.99\substack{+ 0.25\\ -0.20}$  \\
$\nabla_{g-r}$ (9) &  $0.00\substack{+0.00 \\ -0.69}$ &  $-0.52\substack{+ 0.52\\ -0.27}$ &  $0.00\substack{+ 0.00\\ -0.64}$  \\

  \hline
\multicolumn{4}{l}{} \\
 \end{tabular}

\caption*{ \textbf{Notes.}   (1) Stellar mass; (2) star forming rate, in log(M$_{\odot}$/yr); (3) specific star forming rate, in log(yr$^{-1}$); (4) Surface brigthness at the effective radius, in mag/arcsec$^2$; (5) Effective radius in kpc; (6) Sérsic index; (7) axis ratio (b/a); (8) U-V color (9) g-r color gradient (see Equation \ref{eqn:gradi}). The values reported in this table correspond to the median and the upper and lower limits of the central 68 percentile of the distributions. Columns 4 - 7 present the median values across the four ACS filters, which are those used in the clustering algorithm.}

\label{tab:prop_gal_sam}

\end{table}

\subsection{Wavelength dependence of the photometric parameters}

We take advantage of the multi-wavelength data to analyze the possible trends in the photometric parameters with wavelength. We compute whether the median values of each parameter in the sample correlate with the wavelength, as well as for each individual galaxy. Throughout the paper, we calculate the correlation between variables by performing 1000 Monte Carlo and Bootstrap simulations and deriving the Spearman rank coefficient in each one. If more than 95\% of the simulations present a positive (negative) coefficient, we consider that a positive (negative) correlation is present. 

The host galaxy axis ratio $q$ is higher at longer wavelengths, meaning that the galaxies become rounder when seen in the infrared, compared to the UV-optical range. The median value of the sample grows from $q$=0.47 at $F435W$ to $q$=0.67 at $F140W$. For individual galaxies, 46\% of them present a statistically significant positive correlation (the longer the wavelength of the filter considered, the higher the $q$). 

Considering the Sérsic index, the median values of the sample oscillate between $n$=1.0 and $n$=0.8, with no dependencies with wavelength. The only exception is the F435W band, with a lower median value of $n$=0.6.

The median values of the effective radius $R_e$ neither show dependence with wavelength. We find the smallest value in $F435W$ with $R_e$=1.34 kpc, but $F160W$ shows the second smallest one, $R_e$=1.45 kpc. The same applies for individual galaxies.

\cite{Vulcani14} find very clear trends with increasing wavelength, of a $\sim$ 30 \% decrease in size and a $\sim$ 100 \% increase in Sérsic index in our wavelength range. \cite{Lange15} find a less striking relation, with a 16 \% decrease in size from $g$ to $K_s$ band in late-type galaxies. \cite{vanderwel14} find that a similar variation with wavelength, but it fades at lower masses. We use their equation (1) to compute the expected change of effective radius with wavelength.  Assuming the median values in both mass and redshift of our sample, and considering the F850LP and F125W filters, we find $\Delta$log$_{10}($R$_{eff}$[kpc]$)$=0.0098. This is compatible with the measured values of the galaxies in our sample, $\Delta$log$_{10}($R$_{eff}$[kpc]$)$= $0.000\substack{+0.072 \\ -0.046}$.

\subsection{Color gradients of the host galaxies}
\label{sec:colorgrad}
Color gradients have been widely used to understand the structural properties and evolution of galaxies \citep{Franx90,MacArthur04,LaBarbera05,Tortora10,Patel13}. These studies have found negative color gradients, and have been linked to age and metallicity gradients, supporting an inside-out mode of star formation. 

We compute the $g-r$ color gradients of the galaxies along the semi-major axis using the modeled surface brightness in each band. In this way, we consider only the structure of the host galaxy, without contamination due to the star formation regions. We choose the $g$ and $r$ SDSS bands in order to compare with results from the literature. For the K-correction, we follow the same procedure described in Section \ref{colgal}. In addition, we exclude the $F435W$ filter from this analysis since its lower signal-to-noise ratio in the outermost regions of the galaxies prevents us from using it to accurately compute colors. We derive color gradients only in a subsample of galaxies where the segmentation map radius obtained from 3D-HST \citep{Skelton14} is larger than 0.9 arcsecs. We also remove all galaxies where the basic morphological parameters (PA and position of the center) do not match between the two filters used to derive the color. This leaves us with a subsample of 58 galaxies for the study.

We compute the logarithmic slope in bins of 500 pc to derive partial color gradients:
\begin{equation}
\label{eqn:gradi}
$$\nabla_{i,g-r}=\frac{(\mu_g-\mu_r)_{i+1}-(\mu_g-\mu_r)_{i}}{log_{10}(R_{i+1})-log_{10}(R_i)}$$    
\end{equation}
where $R$ is the radius in kpc, and $\mu_g$ and $\mu_r$ are the surface brightness in $g$ and $r$ at that radius respectively. We avoid both the innermost region of the galaxies (R < 500 pc) and regions where more than 30\% of the pixels are covered by the star-formation mask.

\begin{figure}
   \centering
   \includegraphics[width=0.5\textwidth,keepaspectratio]{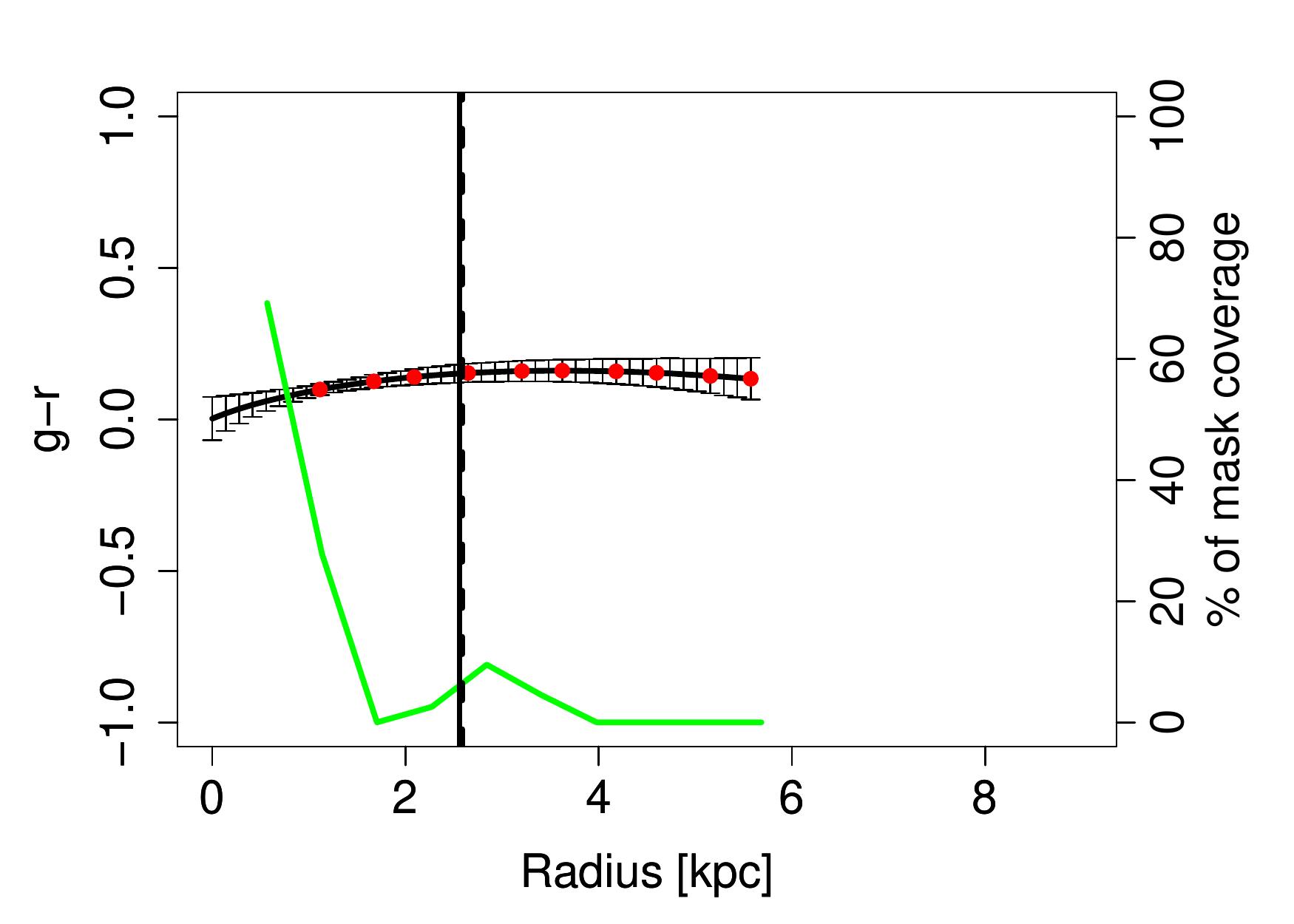}
\includegraphics[width=0.5\textwidth,keepaspectratio]{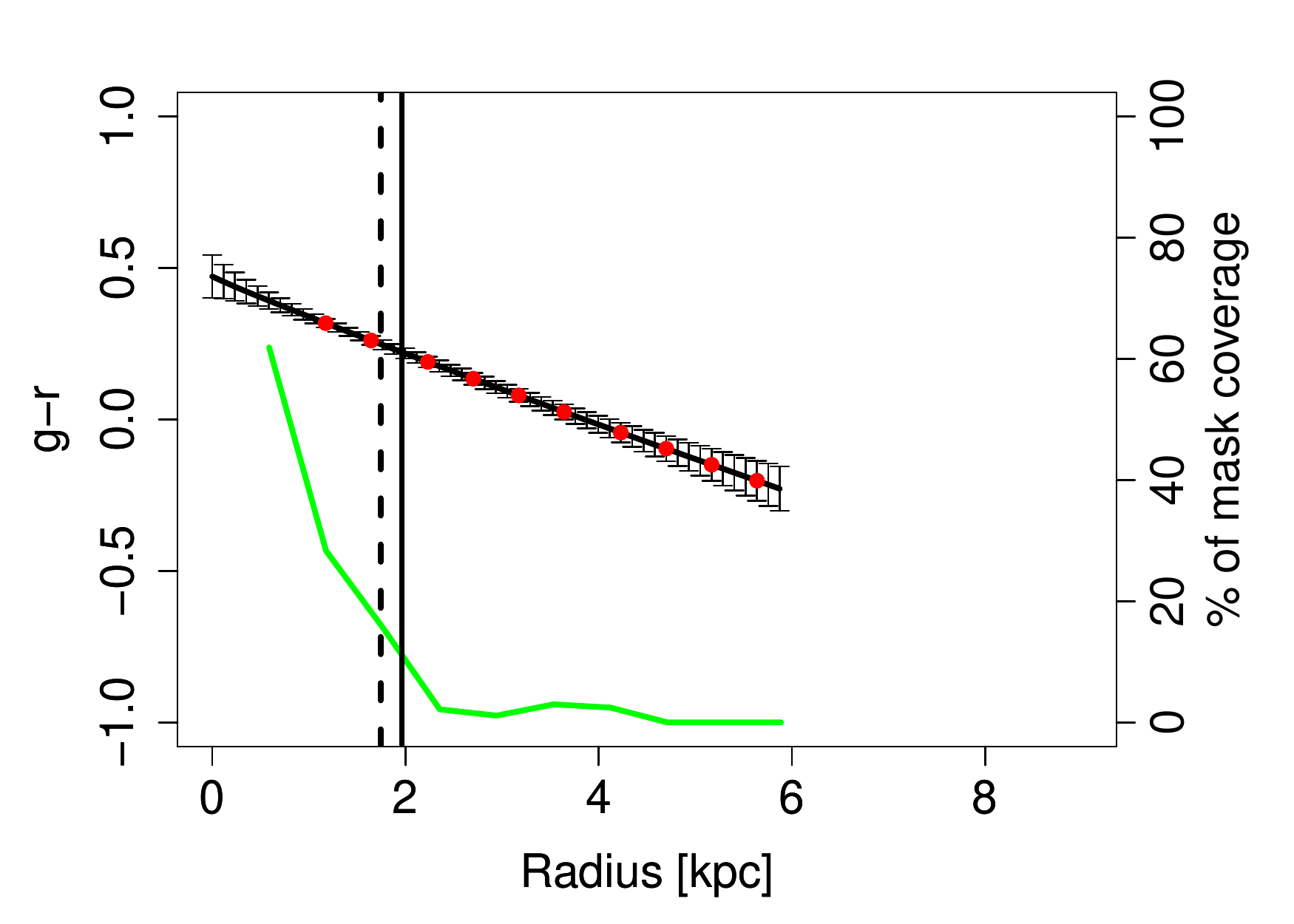}
      \caption{Examples of $g-r$ color profiles for two galaxies in the sample.  \textit{Top.} Galaxy with a flat color profile. \textit{Bottom.} Galaxy with a negative color gradient. The vertical dashed (full) line shows the effective radius in the g (r) band. The green line represents the percentage of the area of the galaxy masked as a function of radius.}
         \label{fig:ex_grad}
   \end{figure}

We present two examples of $g-r$ color as a function of radius in Fig. \ref{fig:ex_grad}, with the first galaxy showing a clear negative gradient and the second, a flat profile. We classify a galaxy with a global gradient when the median value of each partial gradient is larger (in absolute value) than the standard deviation of the whole set of partial gradients. Considering this, we find that only 31 out of the 58 galaxies show a color gradient, with a median value of $\nabla_{g-r}= -0.48$. Out of those 31 galaxies, 28 have a negative color gradient, i.e., the outskirts are bluer than the inner regions. 

{If we consider all 58 galaxies analyzed (assigning $\nabla_{g-r}= 0$ to those where there was no significant gradient) we get a mean value of -0.32. This is comparable to other studies such as \cite{Kennedy15}, where they find $\nabla_{g-r}$ between -0.22 and -0.29 for their galaxies with n<2.5 (as most of our sample is); or to \cite{Tortora10}, who find $\nabla_{g-i} \sim $  -0.2 $\pm$0.15 for late-type galaxies with the same median effective radius as ours.}

\section{Discussion}

In this Section, we analyze the properties of the sample, discovering through clustering analysis the existence of two different classes, according to their radius, Sérsic index, and color. We discuss possible scenarios to explain these differences and possible evolutionary paths between the galaxies in our sample belonging to each class. We also analyze the evidence in favor and against each one of the possibilities.

\subsection{The two classes of host galaxies}
\label{sec:twoclass}
Exploring in detail the distribution of effective radius of our sample of galaxies (Fig. \ref{fig:resul}, panel a), we notice the existence of a double peak, visible in all bands, at $\sim$0.8 kpc and $\sim$1.9 kpc. Selecting galaxies around the two peaks resulted in sub-samples that present a suggestive difference in median values of Sérsic index and U-V color. In order to disentangle more precisely these two possible classes of galaxies, we use a clustering algorithm. We perform the computation using a Gaussian mixture model \citep{GMM}, implemented using the \texttt{sklearn} package in Python \citep{scikit-learn}. This model presupposes that the sample is generated from several Gaussian distributions, and the code estimates their parameters, determining also the cluster to which each galaxy belongs, with a certain confidence level. We assume the sample can be separated into two different clusters (classes A and B) based on the effective radius, the Sérsic index, and the U-V color. For each galaxy, we use the median value across ACS filters for the effective radius and Sérsic index, and the result is presented in Fig.~\ref{fig:clustering}. We proof the existence of two classes, with class A corresponding to the subsample of galaxies with lower radius, higher $n$, and redder U-V color, whereas class B is made up of larger, bluer, and lower $n$ galaxies. Class A consists of 23 galaxies and class B of 72.

We run a series of tests to further confirm and strengthen this conclusion. First, we perform Kolmogorov-Smirnov (KS) tests, which show that classes A and B are statistically different in $n$, R$_e$, and U-V (p-values < 10$^{-4}$ in all cases). In contrast, separating the sample in either $n$, effective radii, or U-V colors instead does not result in statistically different samples in all three parameters simultaneously. The $g-r$ color is also statistically different from one class to the other (redder in class A), giving consistency to our analysis in U-V and $g-r$. {The two subsamples are also present if we run the clustering analysis in each band individually. The KS tests also show that the two subsamples are statistically different in the three parameters ($R_e$, $n$, and U-V) in all filters. Although the subsamples are not identical in all bands, $\sim$ 80\% of the galaxies in each class are present in the majority of filter-selected subsamples.

 \begin{figure*}
   \centering
   \includegraphics[width=0.95\textwidth,keepaspectratio]{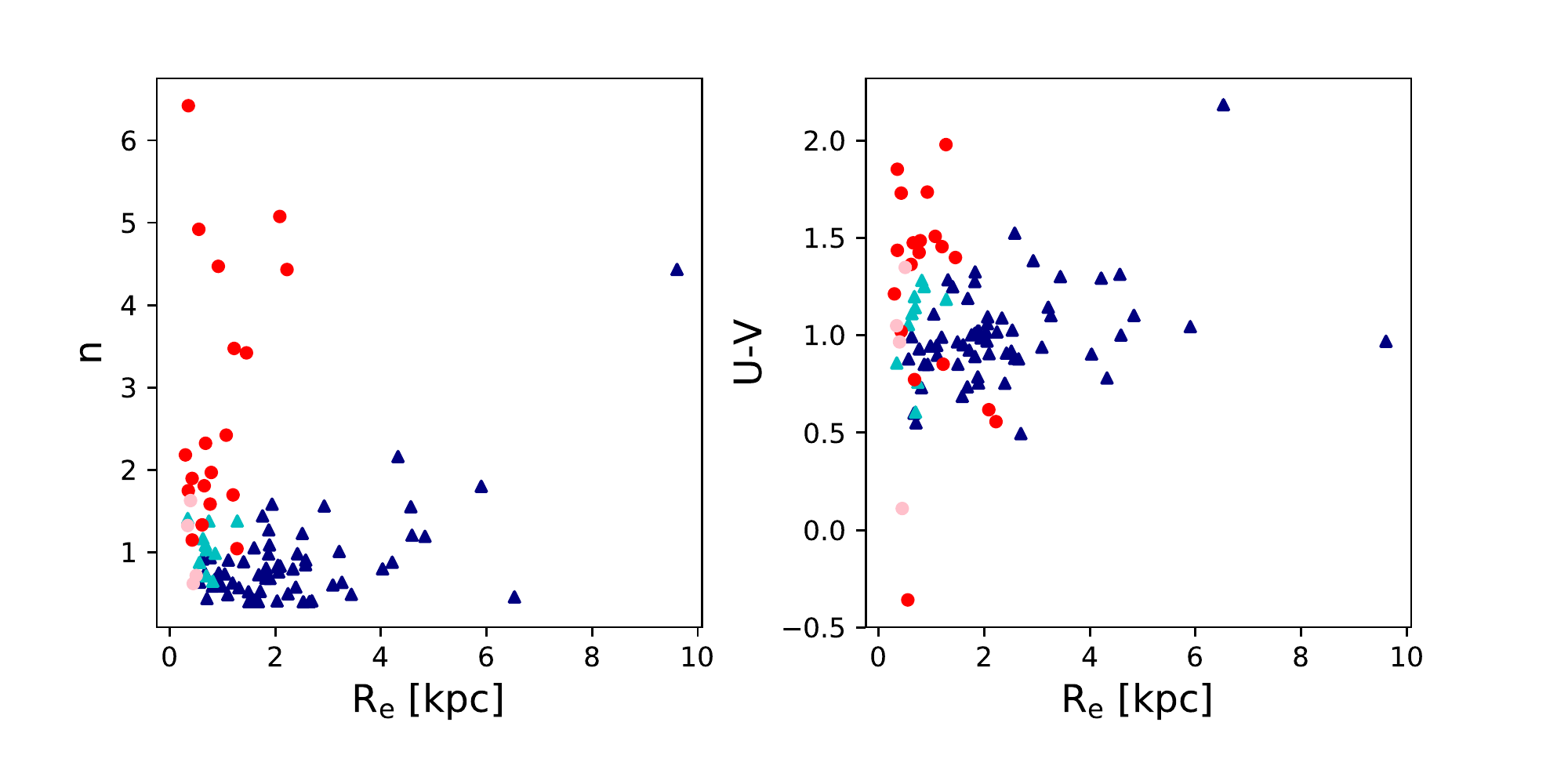}
      \caption{Red and blue dots are galaxies belonging to class A and B with probability p>0.8, respectively. Pink and cyan dots are galaxies that belong to class A and B with $0.8 \geq p>0.5$, respectively. The parameters used in this plot are the median values for each galaxy across all bands.}         
      \label{fig:clustering}
   \end{figure*}

 \begin{figure}
   \centering
   \includegraphics[width=0.5\textwidth,keepaspectratio]{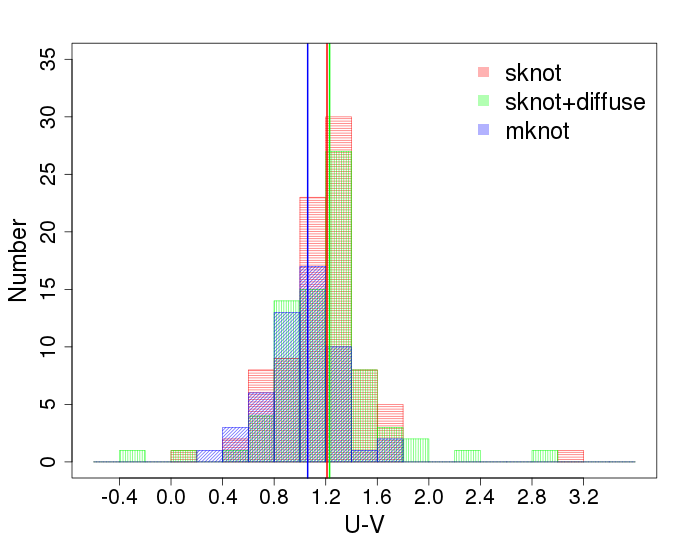}
      \caption{Histogram of U-V color for the starburst galaxies in \cite{Hinojosa-Goni16}. In red, Sknot galaxies, in green Sknot+diffuse, and in blue Mknot. The vertical lines represent the median values of each subsample, with the same color code.  }
         \label{fig:uv_gal_rod}
   \end{figure}

Having confirmed the significance of the two classes, we also analyzed some of the different parameters of each class. We first consider the apparent shape of the galaxies, focusing on the axis ratio $q$. The lowest $q$ value in a sample can be used as a proxy for the intrinsic shape of the galaxies in it, regardless of orientation \citep{Sanchez-Janssen10}. Separating in classes A and B, we find that the minimum $q$ value, in all bands, is lower for class B than for class A, indicating that galaxies in class A are more spheroidal-like than the ones in class B. Furthermore, if we consider the $q_{25\%}$ value (first quartile of the distribution) in each band, the percentage of galaxies in each class that are below that threshold indicates the relative amount of elongated galaxies in each subsample. We find that this percentage is larger in class B than in class A ($\sim$ 31\% versus $\sim$ 18\% in median), strengthening the previous observation of the minimum $q$ values.  We also find that the median specific star forming rate (sSFR) is lower in class A than in class B, with log$_{10}$(sSFR)= $-1.53\substack{+0.94 \\ -0.39}$ log$_{10}$(M$_{\odot}$/yr) and log$_{10}$(sSFR)=$-1.20\substack{+0.60 \\ -0.48}$ log(M$_{\odot}$/yr), respectively. Similarly, the median stellar mass of class A is lower than in class B, with log$_{10}$(M$_{\star}$/M$_{\odot}$)=$8.37\substack{+0.80\\ -0.81}$ and log$_{10}$(M$_{\star}$/M$_{\odot}$)=$8.48\substack{+0.51 \\ -0.79}$, respectively. It is worth noting that in both cases the distributions show significant overlap.

We compare our results with the work by \cite{Hinojosa-Goni16} on starburst galaxies within the COSMOS field. They built their sample using data from Subaru medium band filters \citep{Taniguchi15} in a manner similar to our method in Paper I. Using the HST F814W images, they morphologically classified their sample in three categories: Sknot, when the galaxy consists of a single star-forming clump; Sknot+diffuse, when the single clump is surrounded by diffuse emission; and Mknot, when multiple clumps are identified in a single galaxy. Using their data, we find their Sknot and Sknot+diffuse galaxies to be redder than their Mknot subsample (Fig. \ref{fig:uv_gal_rod}). We applied the K-S test to confirm their distributions are statistically different. The effective radii of their subsamples also show a differentiated behavior, with Mknot galaxies being larger than Sknot+diffuse, which in turn are larger than Sknot. The dispersion in values is, however, very high for the Mknot subsample ($\sim 1.8$ kpc). Both results indicate trends coherent with our findings. This will be explored further in Méndez Abreu et al. 2019 in prep.

Considering the results presented in this section, we find convincing evidence of the existence of two different classes of galaxies hosting star-forming events in our sample: i) larger, bluer, and  disk-like host galaxies, and ii) smaller, generally redder and spheroid-like galaxies.

\subsection{Mass-size relation}

Figure \ref{fig:mass-size} shows the mass-size relation for our sample, with class A galaxies in red and class B in blue. The two classes follow different mass-size relations, with similar slope but a clear offset. The linear fit to each class is shown in red and blue lines, with the 1 $\sigma$ errors of the fit computed using Monte Carlo an Bootstrap simulations and shown as shaded areas. 

The comparison with the literature is also shown in Figure \ref{fig:mass-size}. Note however that our mass range is one or two orders of magnitude lower than theirs. Class B galaxies follow a relation compatible with the extrapolation to low masses of the SFG sample in \cite{vanderwel14}, as well as the disk components of galaxies in the sample studied in \cite{Lange16}. The galaxies in class A, however, follow a relation more similar to the spheroidal and quiescent galaxies.  There is as well an offset in the median values of the effective radius of the two classes, with class A and  showing a median of R$_e$=$0.65 \substack{+0.59 \\ -0.28}$ kpc and $1.82\substack{+1.35 \\ -1.04}$ kpc, respectively.

\begin{figure}
   \centering
   \includegraphics[width=0.5\textwidth,keepaspectratio]{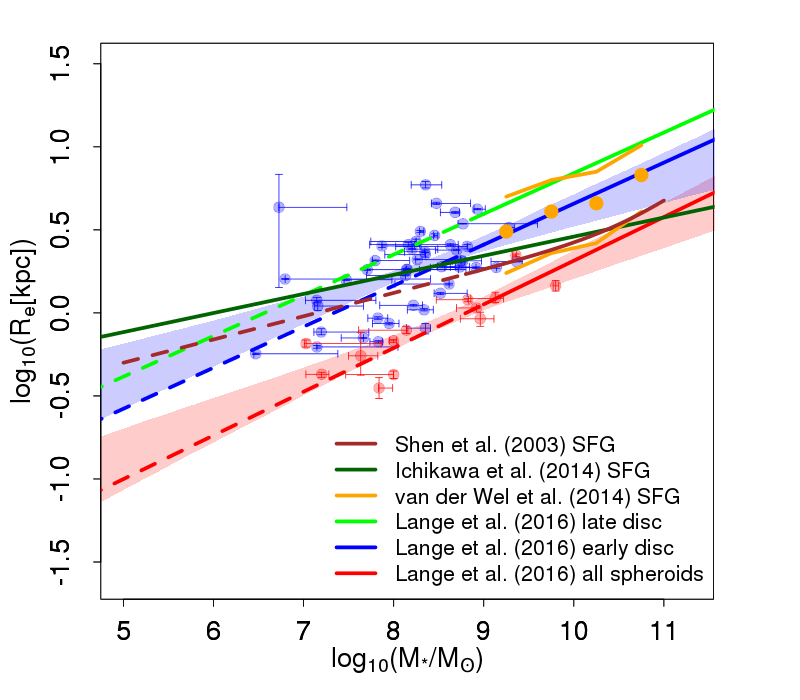}
      \caption{Mass-size relation of our sample of galaxies. Blue dots are galaxies that belong to class B, red dots belong to class A. Shaded blue (red) area are the 1$\sigma$ limits for a linear fit to the blue (red) dots, using bootstrap and Monte Carlo simulations. The lines represent mass-size relations found in the literature (see legend). Lines are continuous until they reach the mass limit for each sample, from where we point extrapolations to lower masses as dashed lines. }
         \label{fig:mass-size}
   \end{figure}

\subsection{Comparison with previous studies on star-forming galaxy subsamples}

In our work, class B galaxies show properties similar to typical SFGs in the literature, while class A present intermediate values between SFG and quiescent galaxies. There are a number of studies in the literature that find a {\it transition} type of galaxies as well, that could be linked with our class A. The different ways of selecting these samples, and the physical characteristics of the transition-type galaxies can provide useful clues to place the results of the present paper in a broader context.

There is a variety of methods for selecting transition galaxies presented in the literature: \cite{Pandya17} selected galaxies below the the main sequence of star formation, \cite{Wang17} identified SFGs in MaNGA with larger 4000 Å break at 1.5$R_e$ than in the center, and \cite{Wang18} selected compact SFGs. Regarding post-starburst galaxies (PSBs), \cite{Pawlik18} (their ePSB class) and \cite{Maltby18} (their low-mass, low redshift PSB) defined transition samples according to principal component analysis (PCA) of their spectra and SED, respectively.

In all those studies where the Sérsic index or the radius were measured, the transition samples presented intermediate values between star-forming and quiescent galaxies in both parameters (in the \cite{Wang18} case, the radius is smaller than in other SFGs by design). In some cases \citep{Pandya17,Pawlik18,Maltby18} the color of the galaxies was measured and presented intermediate values as well. The sSFR shows higher values than in normal SFGs in \cite{Wang18} and \cite{Wang17}, while it is lower by design in \cite{Pandya17}. The metallicities are only measured in \cite{Wang17}, and they are higher than in normal SFGs. In terms of the relative abundance of these transition galaxies, we find 24 \% of the sample to be of the intermediate type. Some other samples are similar to ours (20\% of all the SFG in \citealt{Pandya17}, 14\% in \citealt{Wang17}) while others, specifically the post-starburst samples (\citealt{Maltby18} and \citealt{Pawlik18}) are less common, around 4\%.

Note that our analysis differs from those described before in the way the sub-samples are selected. We do not impose a priori hard limits in some parameters, but analyze the two populations that arise from clustering analysis when considering radius, Sérsic index, and color.

\subsection{Physical explanation for the two-classes subdivision}
In this subsection we review the possible scenarios capable of explaining the existence of two populations of SFGs like the ones presented in this paper. We will discuss processes that entail the evolution of galaxies from class B to class A, but also some models that explain the existence of galaxies which are to stay in class B without necessarily evolving from galaxies in class A.

\subsubsection{Evolution from galaxies in one class to another}

\begin{itemize}
\item \textbf{Major mergers}

If two SFGs (like our class B  population) with similar mass merge, the gas can fall to the center, triggering a fast central starburst. According to \cite{Hopkins10}, the resulting system can be dominated by a central, dense, quasi-spheroidal bulge. If the residual disk is faint, it can be undetectable in our images, and thus the galaxy may have smaller effective radius, while still presenting star-formation, as our class A population \citep[see also][]{Pawlik18}. Dwarf-dwarf galaxy mergers have also been presented as a possible cause of transition-type galaxies, for example in \cite{Dellenbusch08}. However, our galaxies in class A are not systematically more massive than class B  (if anything, the opposite is true, see table \ref{tab:prop_gal_sam}), which would be expected if they were the result of a major merger.

\item \textbf{Violent disk instability}

Cold gas accretion onto a galaxy can produce violent disk instability (VDI), that in turn drives dissipative gas inflows into the disk center (partly as giant star forming clumps coalesce to the center). This results in a compact star forming system ("blue nugget") that is later quenched, turning into a "red nugget" \citep[][among others]{Bournaud07,Ceverino10,Dekel14}. This scenario implies an increase in central surface brightness in the galaxies undergoing this quenching process. We see in Fig. \ref{fig:surf_dens_mass} that the galaxies in our class A subsample are in fact located in the quenched region of the surface mass density vs. mass diagram, considering the relation derived by \cite{Fang13}.

The main limitations to apply this model to our sample is the different mass and redshift range for which it is developed ($M_{\star} > 10^{10} M_{\odot}$, $z=1-3$) compared to our less massive, closer sample ($10^{7} < M_{\star}/M_{\odot}  < 10^{10}$, $z=0.1-0.36$). This affects the availability of cold gas in the intergalactic medium (which is higher at higher redshift) and the ability of our galaxies to accrete it. However, recent works \citep{SanchezAlmeida13,SanchezAlmeida15} present strong evidence of cold flow accretion from the cosmic web into local dwarf galaxies which would make the scenario more plausible. 

\item \textbf{Minor mergers and gas accretion}

Several authors have presented evidence supporting the importance of minor mergers in the star forming history and morphological evolution of galaxies, specifically in recent times ($z$<1) \citep{Bournaud07,Kaviraj14,Smethurst15}. As \cite{Smethurst15} point out, minor mergers could be responsible for the evolution across the green valley of galaxies in an intermediate speed quenching regime ($\sim$ 1.5 Gyr). More specifically, minor mergers would trigger star-formation, that would exhaust the remaining gas and contribute to the quenching of the galaxy \citep{Smethurst15}.

\cite{Starkenburg16b} performed simulations of dwarf galaxies merging with dark satellites ("galaxies" consisting only of dark matter), and found that these interactions triggered starbursts and morphological changes, leaving a more spheroidal and compact system than the original dwarf galaxy. Although this is consistent with our data, it must be noted that dark satellites, even if predicted by cosmological simulations, have not yet been detected.
\end{itemize}

It is worth noting that other scenarios for this evolution (e.g. environment-related processes) cannot be applied to our sample, given the lack of high-density environments in the GOODS-N field.

\subsubsection{No link between the two classes }

Contrary to what has been previously discussed, it is possible that galaxies in different classes just represent different evolutionary paths. 

The existence of a slightly red, small, compact population but with measurable star formation activity (class A) can be explained in several ways. For example, minor merger accretion of gas-rich dwarf galaxies could cause a small level of star formation in an otherwise red, quiescent galaxy. There is evidence of accreted, kinematically misaligned gas in early type and green valley galaxies \citep{Davis11,Jin16}. This could be comparable to the ex-situ clumps of star formation described in \cite{Mandelker14}.

Other authors have recently presented numerical simulations \citep{Wright19,Ledinauskas18} that show reignition of star formation in quiescent, dwarf galaxies due to interactions with streams of gas in the intergalactic medium. This scenario is plausible, considering that the star forming history of local dwarf ellipticals \citep[e.g.][]{Tolstoy09,Grebel98} presents, in some cases, small star forming events at similar redshift as our galaxies (even isolated galaxies as shown in \citealt{Greco18}).

Reaching a conclusion on the precise evolution and nature of the two classes of galaxies cannot be obtained with the data in this paper. However, the similarity with several results in the literature at higher masses, makes us consider that the compaction and quenching scenario is the most likely cause of the two classes. The precise mechanisms that causes it (VDI, minor or major mergers) remain unclear, and all of them may be at play. The exact nature of the compaction, i.e. whether it is a physical contraction with radial migration of the stars, or a relative accumulation of mass in the inner regions compared to the outskirts \citep{Wang18} is also a matter of debate. 

 \begin{figure}
   \centering
   \includegraphics[width=0.5\textwidth,keepaspectratio]{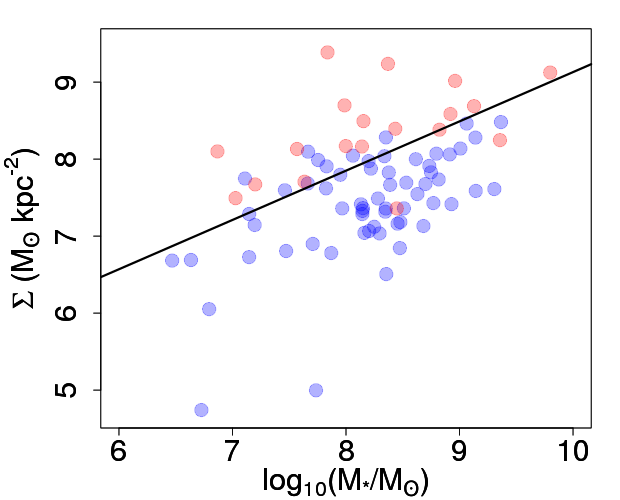}
      \caption{ Surface mass density of the galaxies in the sample as a function of galaxy mass. The line is an extrapolation of the relation presented in Figure 4. of \cite{Fang13} to fit galaxies in the red sequence and the green valley. Red and blue circles represent galaxies in class A and B, respectively.}

         \label{fig:surf_dens_mass}
   \end{figure}

\section{Conclusions}

In this paper we intend to better constrain the relation between morphology of galaxies and their star formation properties. We perform single Sérsic 2D fitting to the HST images of the ELG sample defined in Paper I, using the software \texttt{PHI} \citep{Argyle18}, a Bayesian 2D fitting code, that provides us with realistic uncertainties in the parameters of the fit. 

To accurately measure the light distribution of the underlying host galaxy harboring star-forming events, we need to mask the clumps of star formation. To do so, we create H$\alpha$ images of the galaxies, subtracting a continuum broad-band filter from the filter contaminated by the line, taking into account the slope of the SED. For the galaxies where the H$\alpha$ image does not present clumps, we use the brightest clumps in the F435W images (our bluest filter) as mask. We find that using larger or smaller masks does not change significantly the results, but not using any mask would bias the Sérsic index values in many galaxies (22\% of them would present discrepancies larger than 20\%). 

The sample shows low median Sérsic index ($n \sim$ 1), with 80\% of the sample showing $n$<2 in all bands, consistent with the literature for SFGs. In coherence with the low masses derived in Paper I, the measured radius are small, lower than 3 kpc in at least 80\% of the galaxies. The axis ratio $q$ shows an extended distribution, with growing values of $q$ at longer wavelengths.

We compute also the color of the host component of the galaxies, as well as the color gradients for those galaxies where this analysis is feasible. Most host galaxies belong to the star-forming area of the UVJ diagram, although they are redder than the complete galaxy (considering the star-forming clumps as well). Color gradients are clearly seen in only 31 of the 58 galaxies where we perform the analysis, and the vast majority of them (28) are negative: the outskirts are bluer than the inner regions.

We find, using a clustering algorithm, that the sample can be separated in two classes (A and B) according to their size, Sérsic index, and color. Class A shows lower radius, higher $n$ and redder U-V color, while class B shows the opposite properties. This separation holds in all bands. In addition, the minimum value of the axis ratio $q$ is higher in class A for all bands, indicating that they are more spheroidal-like. In the mass-size diagram, class A galaxies follow a relation similar to passive and spheroidal galaxies. We find several studies in the literature that also describe two different populations of SFGs that are similar to our two classes. 

We review the possible causes of this separation, and find our data most coherent with the 'compaction and quenching' scenario. If this was the case, extended SFGs would become more compact (through mergers or violent disk instability) and spheroidal, and they would later be quenched. This would mean there is an evolutionary link between the galaxies in the two classes. In order to confirm this scenario, a larger sample would be needed. In future work, we will extend this morphological analysis to a sample of higher redshift SFGs.

\begin{acknowledgements}
We thank the anonymous referee whose comments and suggestions helped improve this paper. This work was partly financed by the Spanish Ministerio de Economía y Competitividad (MINECO) within the ESTALLIDOS project (AYA2013-47742-C4-2P and AYA2016-79724-C4-2-P).
AL-C acknowledges financial support from MINECO PhD contract BES-2014-071055 and from grant EEBB-I-16-10913 for a short stay at St. Andrews University.
JMA acknowledge support from the Spanish  Ministerio de Economia y  Competitividad (MINECO) by the grant AYA2017-83204-P.
We thank J. Argyle for help running the code PHI. We also thank R. García-Dias for help with the clustering algorithm and A. Dorta for his work with the computing system CONDOR. AL-C thanks A. S. Borlaff and S. Roca-Fàbrega for helpful discussions. This work has made use of the programming software R \citep{rstat} and the Python package \texttt{astropy} \citep{astropy:2013, astropy:2018}.
\end{acknowledgements}

%-------------------------------------------------------------------
\bibliographystyle{aa}
\bibliography{aa}

\begin{thebibliography}{87}
\expandafter\ifx\csname natexlab\endcsname\relax\def\natexlab#1{#1}\fi

\bibitem[{{Abraham} {et~al.}(1996){Abraham}, {Tanvir}, {Santiago}, {Ellis},
  {Glazebrook}, \& {van den Bergh}}]{Abraham96}
{Abraham}, R.~G., {Tanvir}, N.~R., {Santiago}, B.~X., {et~al.} 1996, \mnras,
  279, L47

\bibitem[{{Amor{\'{\i}}n} {et~al.}(2009){Amor{\'{\i}}n}, {Aguerri},
  {Mu{\~n}oz-Tu{\~n}{\'o}n}, \& {Cair{\'o}s}}]{Amorin09}
{Amor{\'{\i}}n}, R., {Aguerri}, J.~A.~L., {Mu{\~n}oz-Tu{\~n}{\'o}n}, C., \&
  {Cair{\'o}s}, L.~M. 2009, \aap, 501, 75

\bibitem[{{Amor{\'{\i}}n} {et~al.}(2012){Amor{\'{\i}}n}, {P{\'e}rez-Montero},
  {V{\'{\i}}lchez}, \& {Papaderos}}]{Amorin12}
{Amor{\'{\i}}n}, R., {P{\'e}rez-Montero}, E., {V{\'{\i}}lchez}, J.~M., \&
  {Papaderos}, P. 2012, \apj, 749, 185

\bibitem[{{Argyle} {et~al.}(2018){Argyle}, {M{\'e}ndez-Abreu}, {Wild}, \&
  {Mortlock}}]{Argyle18}
{Argyle}, J.~J., {M{\'e}ndez-Abreu}, J., {Wild}, V., \& {Mortlock}, D.~J. 2018,
  \mnras, 479, 3076

\bibitem[{{Astropy Collaboration} {et~al.}(2013){Astropy Collaboration},
  {Robitaille}, {Tollerud}, {Greenfield}, {Droettboom}, {Bray}, {Aldcroft},
  {Davis}, {Ginsburg}, {Price-Whelan}, {Kerzendorf}, {Conley}, {Crighton},
  {Barbary}, {Muna}, {Ferguson}, {Grollier}, {Parikh}, {Nair}, {Unther},
  {Deil}, {Woillez}, {Conseil}, {Kramer}, {Turner}, {Singer}, {Fox}, {Weaver},
  {Zabalza}, {Edwards}, {Azalee Bostroem}, {Burke}, {Casey}, {Crawford},
  {Dencheva}, {Ely}, {Jenness}, {Labrie}, {Lim}, {Pierfederici}, {Pontzen},
  {Ptak}, {Refsdal}, {Servillat}, \& {Streicher}}]{astropy:2013}
{Astropy Collaboration}, {Robitaille}, T.~P., {Tollerud}, E.~J., {et~al.} 2013,
  \aap, 558, A33

\bibitem[{{Bauer} {et~al.}(2013){Bauer}, {Hopkins}, {Gunawardhana}, {Taylor},
  {Baldry}, {Bamford}, {Bland-Hawthorn}, {Brough}, {Brown}, {Cluver},
  {Colless}, {Conselice}, {Croom}, {Driver}, {Foster}, {Jones}, {Lara-Lopez},
  {Liske}, {L{\'o}pez-S{\'a}nchez}, {Loveday}, {Norberg}, {Owers}, {Pimbblet},
  {Robotham}, {Sansom}, \& {Sharp}}]{Bauer13}
{Bauer}, A.~E., {Hopkins}, A.~M., {Gunawardhana}, M., {et~al.} 2013, \mnras,
  434, 209

\bibitem[{{Bertin} \& {Arnouts}(1996)}]{Bertin96}
{Bertin}, E. \& {Arnouts}, S. 1996, \aaps, 117, 393

\bibitem[{{Bournaud} {et~al.}(2007){Bournaud}, {Elmegreen}, \&
  {Elmegreen}}]{Bournaud07}
{Bournaud}, F., {Elmegreen}, B.~G., \& {Elmegreen}, D.~M. 2007, \apj, 670, 237

\bibitem[{Bouveyron {et~al.}(2007)Bouveyron, Girard, \& Schmid}]{GMM}
Bouveyron, C., Girard, S., \& Schmid, C. 2007, Computational Statistics \& Data
  Analysis, 52, 502

\bibitem[{{Brammer} {et~al.}(2012){Brammer}, {van Dokkum}, {Franx},
  {Fumagalli}, {Patel}, {Rix}, {Skelton}, {Kriek}, {Nelson}, {Schmidt},
  {Bezanson}, {da Cunha}, {Erb}, {Fan}, {F{\"o}rster Schreiber}, {Illingworth},
  {Labb{\'e}}, {Leja}, {Lundgren}, {Magee}, {Marchesini}, {McCarthy},
  {Momcheva}, {Muzzin}, {Quadri}, {Steidel}, {Tal}, {Wake}, {Whitaker}, \&
  {Williams}}]{Brammer12}
{Brammer}, G.~B., {van Dokkum}, P.~G., {Franx}, M., {et~al.} 2012, \apjs, 200,
  13

\bibitem[{{Brennan} {et~al.}(2017){Brennan}, {Pandya}, {Somerville}, {Barro},
  {Bluck}, {Taylor}, {Wuyts}, {Bell}, {Dekel}, {Faber}, {Ferguson},
  {Koekemoer}, {Kurczynski}, {McIntosh}, {Newman}, \& {Primack}}]{Brennan17}
{Brennan}, R., {Pandya}, V., {Somerville}, R.~S., {et~al.} 2017, \mnras, 465,
  619

\bibitem[{{Cair{\'o}s} {et~al.}(2002){Cair{\'o}s}, {Caon},
  {Garc{\'{\i}}a-Lorenzo}, {V{\'{\i}}lchez}, \&
  {Mu{\~n}oz-Tu{\~n}{\'o}n}}]{Cairos02}
{Cair{\'o}s}, L.~M., {Caon}, N., {Garc{\'{\i}}a-Lorenzo}, B., {V{\'{\i}}lchez},
  J.~M., \& {Mu{\~n}oz-Tu{\~n}{\'o}n}, C. 2002, \apj, 577, 164

\bibitem[{{Cair{\'o}s} {et~al.}(2003){Cair{\'o}s}, {Caon}, {Papaderos},
  {Noeske}, {V{\'{\i}}lchez}, {Garc{\'{\i}}a Lorenzo}, \&
  {Mu{\~n}oz-Tu{\~n}{\'o}n}}]{Cairos03}
{Cair{\'o}s}, L.~M., {Caon}, N., {Papaderos}, P., {et~al.} 2003, \apj, 593, 312

\bibitem[{{Caon} {et~al.}(2005){Caon}, {Cair{\'o}s}, {Aguerri}, \&
  {Mu{\~n}oz-Tu{\~n}{\'o}n}}]{Caon05}
{Caon}, N., {Cair{\'o}s}, L.~M., {Aguerri}, J.~A.~L., \&
  {Mu{\~n}oz-Tu{\~n}{\'o}n}, C. 2005, \apjs, 157, 218

\bibitem[{{Cava} {et~al.}(2018){Cava}, {Schaerer}, {Richard},
  {P{\'e}rez-Gonz{\'a}lez}, {Dessauges-Zavadsky}, {Mayer}, \&
  {Tamburello}}]{Cava18}
{Cava}, A., {Schaerer}, D., {Richard}, J., {et~al.} 2018, Nature Astronomy, 2,
  76

\bibitem[{{Cepa} {et~al.}(2000){Cepa}, {Aguiar}, {Escalera},
  {Gonzalez-Serrano}, {Joven-Alvarez}, {Peraza}, {Rasilla}, {Rodriguez-Ramos},
  {Gonzalez}, {Cobos Duenas}, {Sanchez}, {Tejada}, {Bland-Hawthorn},
  {Militello}, \& {Rosa}}]{Cepa00}
{Cepa}, J., {Aguiar}, M., {Escalera}, V.~G., {et~al.} 2000, in \procspie, Vol.
  4008, Optical and IR Telescope Instrumentation and Detectors, ed. M.~{Iye} \&
  A.~F. {Moorwood}, 623--631

\bibitem[{{Ceverino} {et~al.}(2010){Ceverino}, {Dekel}, \&
  {Bournaud}}]{Ceverino10}
{Ceverino}, D., {Dekel}, A., \& {Bournaud}, F. 2010, \mnras, 404, 2151

\bibitem[{{Davis} {et~al.}(2011){Davis}, {Alatalo}, {Sarzi}, {Bureau}, {Young},
  {Blitz}, {Serra}, {Crocker}, {Krajnovi{\'c}}, {McDermid}, {Bois}, {Bournaud},
  {Cappellari}, {Davies}, {Duc}, {de Zeeuw}, {Emsellem}, {Khochfar},
  {Kuntschner}, {Lablanche}, {Morganti}, {Naab}, {Oosterloo}, {Scott}, \&
  {Weijmans}}]{Davis11}
{Davis}, T.~A., {Alatalo}, K., {Sarzi}, M., {et~al.} 2011, \mnras, 417, 882

\bibitem[{{de Souza} {et~al.}(2004){de Souza}, {Gadotti}, \& {dos
  Anjos}}]{deSouza04}
{de Souza}, R.~E., {Gadotti}, D.~A., \& {dos Anjos}, S. 2004, \apjs, 153, 411

\bibitem[{{Dekel} \& {Burkert}(2014)}]{Dekel14}
{Dekel}, A. \& {Burkert}, A. 2014, \mnras, 438, 1870

\bibitem[{{Dellenbusch} {et~al.}(2008){Dellenbusch}, {Gallagher}, {Knezek}, \&
  {Noble}}]{Dellenbusch08}
{Dellenbusch}, K.~E., {Gallagher}, III, J.~S., {Knezek}, P.~M., \& {Noble},
  A.~G. 2008, \aj, 135, 326

\bibitem[{{Di Matteo} {et~al.}(2008){Di Matteo}, {Bournaud}, {Martig},
  {Combes}, {Melchior}, \& {Semelin}}]{DiMatteo08}
{Di Matteo}, P., {Bournaud}, F., {Martig}, M., {et~al.} 2008, \aap, 492, 31

\bibitem[{{Elmegreen} {et~al.}(2004{\natexlab{a}}){Elmegreen}, {Elmegreen}, \&
  {Hirst}}]{Elmegreen04b}
{Elmegreen}, D.~M., {Elmegreen}, B.~G., \& {Hirst}, A.~C. 2004{\natexlab{a}},
  \apjl, 604, L21

\bibitem[{{Elmegreen} {et~al.}(2007){Elmegreen}, {Elmegreen}, {Ravindranath},
  \& {Coe}}]{Elmegreen07}
{Elmegreen}, D.~M., {Elmegreen}, B.~G., {Ravindranath}, S., \& {Coe}, D.~A.
  2007, \apj, 658, 763

\bibitem[{{Elmegreen} {et~al.}(2004{\natexlab{b}}){Elmegreen}, {Elmegreen}, \&
  {Sheets}}]{Elmegreen04}
{Elmegreen}, D.~M., {Elmegreen}, B.~G., \& {Sheets}, C.~M. 2004{\natexlab{b}},
  \apj, 603, 74

\bibitem[{{Fang} {et~al.}(2013){Fang}, {Faber}, {Koo}, \& {Dekel}}]{Fang13}
{Fang}, J.~J., {Faber}, S.~M., {Koo}, D.~C., \& {Dekel}, A. 2013, \apj, 776, 63

\bibitem[{{Faucher-Gigu{\`e}re}(2018)}]{Faucher-guiguere18}
{Faucher-Gigu{\`e}re}, C.-A. 2018, \mnras, 473, 3717

\bibitem[{{Fillingham} {et~al.}(2018){Fillingham}, {Cooper}, {Boylan-Kolchin},
  {Bullock}, {Garrison-Kimmel}, \& {Wheeler}}]{Fillingham18}
{Fillingham}, S.~P., {Cooper}, M.~C., {Boylan-Kolchin}, M., {et~al.} 2018,
  \mnras, 477, 4491

\bibitem[{{Franx} \& {Illingworth}(1990)}]{Franx90}
{Franx}, M. \& {Illingworth}, G. 1990, \apjl, 359, L41

\bibitem[{{Grebel}(1998)}]{Grebel98}
{Grebel}, E.~K. 1998, Highlights of Astronomy, 11, 125

\bibitem[{{Greco} {et~al.}(2018){Greco}, {Goulding}, {Greene}, {Strauss},
  {Huang}, {Kim}, \& {Komiyama}}]{Greco18}
{Greco}, J.~P., {Goulding}, A.~D., {Greene}, J.~E., {et~al.} 2018, \apj, 866,
  112

\bibitem[{{Grogin} {et~al.}(2011){Grogin}, {Kocevski}, {Faber}, {Ferguson},
  {Koekemoer}, {Riess}, {Acquaviva}, {Alexander}, {Almaini}, {Ashby}, {Barden},
  {Bell}, {Bournaud}, {Brown}, {Caputi}, {Casertano}, {Cassata}, {Castellano},
  {Challis}, {Chary}, {Cheung}, {Cirasuolo}, {Conselice}, {Roshan Cooray},
  {Croton}, {Daddi}, {Dahlen}, {Dav{\'e}}, {de Mello}, {Dekel}, {Dickinson},
  {Dolch}, {Donley}, {Dunlop}, {Dutton}, {Elbaz}, {Fazio}, {Filippenko},
  {Finkelstein}, {Fontana}, {Gardner}, {Garnavich}, {Gawiser}, {Giavalisco},
  {Grazian}, {Guo}, {Hathi}, {H{\"a}ussler}, {Hopkins}, {Huang}, {Huang},
  {Jha}, {Kartaltepe}, {Kirshner}, {Koo}, {Lai}, {Lee}, {Li}, {Lotz}, {Lucas},
  {Madau}, {McCarthy}, {McGrath}, {McIntosh}, {McLure}, {Mobasher},
  {Moustakas}, {Mozena}, {Nandra}, {Newman}, {Niemi}, {Noeske}, {Papovich},
  {Pentericci}, {Pope}, {Primack}, {Rajan}, {Ravindranath}, {Reddy}, {Renzini},
  {Rix}, {Robaina}, {Rodney}, {Rosario}, {Rosati}, {Salimbeni}, {Scarlata},
  {Siana}, {Simard}, {Smidt}, {Somerville}, {Spinrad}, {Straughn}, {Strolger},
  {Telford}, {Teplitz}, {Trump}, {van der Wel}, {Villforth}, {Wechsler},
  {Weiner}, {Wiklind}, {Wild}, {Wilson}, {Wuyts}, {Yan}, \& {Yun}}]{Grogin11}
{Grogin}, N.~A., {Kocevski}, D.~D., {Faber}, S.~M., {et~al.} 2011, \apjs, 197,
  35

\bibitem[{{Hinojosa-Go{\~n}i} {et~al.}(2016){Hinojosa-Go{\~n}i},
  {Mu{\~n}oz-Tu{\~n}{\'o}n}, \& {M{\'e}ndez-Abreu}}]{Hinojosa-Goni16}
{Hinojosa-Go{\~n}i}, R., {Mu{\~n}oz-Tu{\~n}{\'o}n}, C., \& {M{\'e}ndez-Abreu},
  J. 2016, \aap, 592, A122

\bibitem[{{Hogg} {et~al.}(2002){Hogg}, {Baldry}, {Blanton}, \&
  {Eisenstein}}]{Hogg02}
{Hogg}, D.~W., {Baldry}, I.~K., {Blanton}, M.~R., \& {Eisenstein}, D.~J. 2002,
  arXiv Astrophysics e-prints

\bibitem[{{Hopkins} {et~al.}(2010){Hopkins}, {Croton}, {Bundy}, {Khochfar},
  {van den Bosch}, {Somerville}, {Wetzel}, {Keres}, {Hernquist}, {Stewart},
  {Younger}, {Genel}, \& {Ma}}]{Hopkins10}
{Hopkins}, P.~F., {Croton}, D., {Bundy}, K., {et~al.} 2010, \apj, 724, 915

\bibitem[{{Ilbert} {et~al.}(2013){Ilbert}, {McCracken}, {Le F{\`e}vre},
  {Capak}, {Dunlop}, {Karim}, {Renzini}, {Caputi}, {Boissier}, {Arnouts},
  {Aussel}, {Comparat}, {Guo}, {Hudelot}, {Kartaltepe}, {Kneib}, {Krogager},
  {Le Floc'h}, {Lilly}, {Mellier}, {Milvang-Jensen}, {Moutard}, {Onodera},
  {Richard}, {Salvato}, {Sanders}, {Scoville}, {Silverman}, {Taniguchi},
  {Tasca}, {Thomas}, {Toft}, {Tresse}, {Vergani}, {Wolk}, \& {Zirm}}]{Ilbert13}
{Ilbert}, O., {McCracken}, H.~J., {Le F{\`e}vre}, O., {et~al.} 2013, \aap, 556,
  A55

\bibitem[{{Janowiecki} \& {Salzer}(2014)}]{Janowiecki14}
{Janowiecki}, S. \& {Salzer}, J.~J. 2014, \apj, 793, 109

\bibitem[{{Jin} {et~al.}(2016){Jin}, {Chen}, {Shi}, {Tremonti}, {Bershady},
  {Merrifield}, {Emsellem}, {Fu}, {Wake}, {Bundy}, {Lin}, {Argudo-Fernandez},
  {Huang}, {Stark}, {Storchi-Bergmann}, {Bizyaev}, {Brownstein}, {Chisholm},
  {Guo}, {Hao}, {Hu}, {Li}, {Li}, {Masters}, {Malanushenko}, {Pan}, {Riffel},
  {Roman-Lopes}, {Simmons}, {Thomas}, {Wang}, {Westfall}, \& {Yan}}]{Jin16}
{Jin}, Y., {Chen}, Y., {Shi}, Y., {et~al.} 2016, \mnras, 463, 913

\bibitem[{{Kauffmann} {et~al.}(2003){Kauffmann}, {Heckman}, {White}, {Charlot},
  {Tremonti}, {Peng}, {Seibert}, {Brinkmann}, {Nichol}, {SubbaRao}, \&
  {York}}]{Kauffmann03}
{Kauffmann}, G., {Heckman}, T.~M., {White}, S.~D.~M., {et~al.} 2003, \mnras,
  341, 54

\bibitem[{{Kaviraj}(2014)}]{Kaviraj14}
{Kaviraj}, S. 2014, \mnras, 440, 2944

\bibitem[{{Kelvin} {et~al.}(2014){Kelvin}, {Driver}, {Robotham}, {Graham},
  {Phillipps}, {Agius}, {Alpaslan}, {Baldry}, {Bamford}, {Bland-Hawthorn},
  {Brough}, {Brown}, {Colless}, {Conselice}, {Hopkins}, {Liske}, {Loveday},
  {Norberg}, {Pimbblet}, {Popescu}, {Prescott}, {Taylor}, \&
  {Tuffs}}]{Kelvin14}
{Kelvin}, L.~S., {Driver}, S.~P., {Robotham}, A.~S.~G., {et~al.} 2014, \mnras,
  439, 1245

\bibitem[{{Kennedy} {et~al.}(2015){Kennedy}, {Bamford}, {Baldry},
  {H{\"a}u{\ss}ler}, {Holwerda}, {Hopkins}, {Kelvin}, {Lange}, {Moffett},
  {Popescu}, {Taylor}, {Tuffs}, {Vika}, \& {Vulcani}}]{Kennedy15}
{Kennedy}, R., {Bamford}, S.~P., {Baldry}, I., {et~al.} 2015, \mnras, 454, 806

\bibitem[{{Koleva} {et~al.}(2013){Koleva}, {Bouchard}, {Prugniel}, {De Rijcke},
  \& {Vauglin}}]{Koleva13}
{Koleva}, M., {Bouchard}, A., {Prugniel}, P., {De Rijcke}, S., \& {Vauglin}, I.
  2013, \mnras, 428, 2949

\bibitem[{{La Barbera} {et~al.}(2005){La Barbera}, {de Carvalho}, {Gal},
  {Busarello}, {Merluzzi}, {Capaccioli}, \& {Djorgovski}}]{LaBarbera05}
{La Barbera}, F., {de Carvalho}, R.~R., {Gal}, R.~R., {et~al.} 2005, \apjl,
  626, L19

\bibitem[{{Lange} {et~al.}(2015){Lange}, {Driver}, {Robotham}, {Kelvin},
  {Graham}, {Alpaslan}, {Andrews}, {Baldry}, {Bamford}, {Bland-Hawthorn},
  {Brough}, {Cluver}, {Conselice}, {Davies}, {Haeussler}, {Konstantopoulos},
  {Loveday}, {Moffett}, {Norberg}, {Phillipps}, {Taylor},
  {L{\'o}pez-S{\'a}nchez}, \& {Wilkins}}]{Lange15}
{Lange}, R., {Driver}, S.~P., {Robotham}, A.~S.~G., {et~al.} 2015, \mnras, 447,
  2603

\bibitem[{{Lange} {et~al.}(2016){Lange}, {Moffett}, {Driver}, {Robotham},
  {Lagos}, {Kelvin}, {Conselice}, {Margalef-Bentabol}, {Alpaslan}, {Baldry},
  {Bland-Hawthorn}, {Bremer}, {Brough}, {Cluver}, {Colless}, {Davies},
  {H{\"a}u{\ss}ler}, {Holwerda}, {Hopkins}, {Kafle}, {Kennedy}, {Liske},
  {Phillipps}, {Popescu}, {Taylor}, {Tuffs}, {van Kampen}, \&
  {Wright}}]{Lange16}
{Lange}, R., {Moffett}, A.~J., {Driver}, S.~P., {et~al.} 2016, \mnras, 462,
  1470

\bibitem[{{Ledinauskas} \& {Zubovas}(2018)}]{Ledinauskas18}
{Ledinauskas}, E. \& {Zubovas}, K. 2018, \aap, 615, A64

\bibitem[{{Lian} {et~al.}(2015){Lian}, {Kong}, {Jiang}, {Yan}, \&
  {Gao}}]{Lian15}
{Lian}, J.~H., {Kong}, X., {Jiang}, N., {Yan}, W., \& {Gao}, Y.~L. 2015,
  \mnras, 451, 1130

\bibitem[{{Lumbreras-Calle} {et~al.}(2019){Lumbreras-Calle},
  {Mu{\~n}oz-Tu{\~n}{\'o}n}, {M{\'e}ndez-Abreu}, {Mas-Hesse},
  {P{\'e}rez-Gonz{\'a}lez}, {Alcalde Pampliega}, {Arrabal Haro}, {Cava},
  {Dom{\'{\i}}nguez S{\'a}nchez}, {Eliche-Moral}, {Alonso-Herrero}, {Borlaff},
  {Gallego}, {Hern{\'a}n-Caballero}, {Koekemoer}, \&
  {Rodr{\'{\i}}guez-Mu{\~n}oz}}]{Lumbreras-Calle18}
{Lumbreras-Calle}, A., {Mu{\~n}oz-Tu{\~n}{\'o}n}, C., {M{\'e}ndez-Abreu}, J.,
  {et~al.} 2019, \aap, 621, A52

\bibitem[{{MacArthur} {et~al.}(2004){MacArthur}, {Courteau}, {Bell}, \&
  {Holtzman}}]{MacArthur04}
{MacArthur}, L.~A., {Courteau}, S., {Bell}, E., \& {Holtzman}, J.~A. 2004,
  \apjs, 152, 175

\bibitem[{{Maltby} {et~al.}(2018){Maltby}, {Almaini}, {Wild}, {Hatch},
  {Hartley}, {Simpson}, {Rowlands}, \& {Socolovsky}}]{Maltby18}
{Maltby}, D.~T., {Almaini}, O., {Wild}, V., {et~al.} 2018, \mnras, 480, 381

\bibitem[{{Mandelker} {et~al.}(2014){Mandelker}, {Dekel}, {Ceverino}, {Tweed},
  {Moody}, \& {Primack}}]{Mandelker14}
{Mandelker}, N., {Dekel}, A., {Ceverino}, D., {et~al.} 2014, \mnras, 443, 3675

\bibitem[{{Martig} {et~al.}(2009){Martig}, {Bournaud}, {Teyssier}, \&
  {Dekel}}]{Martig09}
{Martig}, M., {Bournaud}, F., {Teyssier}, R., \& {Dekel}, A. 2009, \apj, 707,
  250

\bibitem[{{M{\'e}ndez-Abreu} {et~al.}(2008){M{\'e}ndez-Abreu}, {Aguerri},
  {Corsini}, \& {Simonneau}}]{Mendez-Abreu08}
{M{\'e}ndez-Abreu}, J., {Aguerri}, J.~A.~L., {Corsini}, E.~M., \& {Simonneau},
  E. 2008, \aap, 478, 353

\bibitem[{{Meyer} {et~al.}(2014){Meyer}, {Lisker}, {Janz}, \&
  {Papaderos}}]{Meyer14}
{Meyer}, H.~T., {Lisker}, T., {Janz}, J., \& {Papaderos}, P. 2014, \aap, 562,
  A49

\bibitem[{{Oliver} {et~al.}(2010){Oliver}, {Frost}, {Farrah},
  {Gonzalez-Solares}, {Shupe}, {Henriques}, {Roseboom}, {Alfonso-Luis},
  {Babbedge}, {Frayer}, {Lencz}, {Lonsdale}, {Masci}, {Padgett}, {Polletta},
  {Rowan-Robinson}, {Siana}, {Smith}, {Surace}, \& {Vaccari}}]{Oliver10}
{Oliver}, S., {Frost}, M., {Farrah}, D., {et~al.} 2010, \mnras, 405, 2279

\bibitem[{{Pandya} {et~al.}(2017){Pandya}, {Brennan}, {Somerville}, {Choi},
  {Barro}, {Wuyts}, {Taylor}, {Behroozi}, {Kirkpatrick}, {Faber}, {Primack},
  {Koo}, {McIntosh}, {Kocevski}, {Bell}, {Dekel}, {Fang}, {Ferguson}, {Grogin},
  {Koekemoer}, {Lu}, {Mantha}, {Mobasher}, {Newman}, {Pacifici}, {Papovich},
  {van der Wel}, \& {Yesuf}}]{Pandya17}
{Pandya}, V., {Brennan}, R., {Somerville}, R.~S., {et~al.} 2017, \mnras, 472,
  2054

\bibitem[{{Patel} {et~al.}(2013){Patel}, {van Dokkum}, {Franx}, {Quadri},
  {Muzzin}, {Marchesini}, {Williams}, {Holden}, \& {Stefanon}}]{Patel13}
{Patel}, S.~G., {van Dokkum}, P.~G., {Franx}, M., {et~al.} 2013, \apj, 766, 15

\bibitem[{{Paulino-Afonso} {et~al.}(2017){Paulino-Afonso}, {Sobral},
  {Buitrago}, \& {Afonso}}]{Paulino-Afonso17}
{Paulino-Afonso}, A., {Sobral}, D., {Buitrago}, F., \& {Afonso}, J. 2017,
  \mnras, 465, 2717

\bibitem[{{Pawlik} {et~al.}(2018){Pawlik}, {Taj Aldeen}, {Wild},
  {Mendez-Abreu}, {Lah{\'e}n}, {Johansson}, {Jimenez}, {Lucas}, {Zheng},
  {Walcher}, \& {Rowlands}}]{Pawlik18}
{Pawlik}, M.~M., {Taj Aldeen}, L., {Wild}, V., {et~al.} 2018, \mnras, 477, 1708

\bibitem[{Pedregosa {et~al.}(2011)Pedregosa, Varoquaux, Gramfort, Michel,
  Thirion, Grisel, Blondel, Prettenhofer, Weiss, Dubourg, Vanderplas, Passos,
  Cournapeau, Brucher, Perrot, \& Duchesnay}]{scikit-learn}
Pedregosa, F., Varoquaux, G., Gramfort, A., {et~al.} 2011, Journal of Machine
  Learning Research, 12, 2825

\bibitem[{{Peng} {et~al.}(2002){Peng}, {Ho}, {Impey}, \& {Rix}}]{Peng02}
{Peng}, C.~Y., {Ho}, L.~C., {Impey}, C.~D., \& {Rix}, H.-W. 2002, \aj, 124, 266

\bibitem[{{P{\'e}rez-Gonz{\'a}lez} {et~al.}(2013){P{\'e}rez-Gonz{\'a}lez},
  {Cava}, {Barro}, {Villar}, {Cardiel}, {Ferreras},
  {Rodr{\'{\i}}guez-Espinosa}, {Alonso-Herrero}, {Balcells}, {Cenarro}, {Cepa},
  {Charlot}, {Cimatti}, {Conselice}, {Daddi}, {Donley}, {Elbaz}, {Espino},
  {Gallego}, {Gobat}, {Gonz{\'a}lez-Mart{\'{\i}}n}, {Guzm{\'a}n},
  {Hern{\'a}n-Caballero}, {Mu{\~n}oz-Tu{\~n}{\'o}n}, {Renzini},
  {Rodr{\'{\i}}guez-Zaur{\'{\i}}n}, {Tresse}, {Trujillo}, \&
  {Zamorano}}]{Perez-Gonzalez13}
{P{\'e}rez-Gonz{\'a}lez}, P.~G., {Cava}, A., {Barro}, G., {et~al.} 2013, \apj,
  762, 46

\bibitem[{{Price-Whelan} {et~al.}(2018){Price-Whelan}, {Sip{\H{o}}cz},
  {G{\"u}nther}, {Lim}, {Crawford}, {Conseil}, {Shupe}, {Craig}, {Dencheva},
  {Ginsburg}, {VanderPlas}, {Bradley}, {P{\'e}rez-Su{\'a}rez}, {de Val-Borro},
  {Paper Contributors}, {Aldcroft}, {Cruz}, {Robitaille}, {Tollerud},
  {Coordination Committee}, {Ardelean}, {Babej}, {Bach}, {Bachetti}, {Bakanov},
  {Bamford}, {Barentsen}, {Barmby}, {Baumbach}, {Berry}, {Biscani}, {Boquien},
  {Bostroem}, {Bouma}, {Brammer}, {Bray}, {Breytenbach}, {Buddelmeijer},
  {Burke}, {Calderone}, {Cano Rodr{\'\i}guez}, {Cara}, {Cardoso}, {Cheedella},
  {Copin}, {Corrales}, {Crichton}, {D{\textquoteright}Avella}, {Deil},
  {Depagne}, {Dietrich}, {Donath}, {Droettboom}, {Earl}, {Erben}, {Fabbro},
  {Ferreira}, {Finethy}, {Fox}, {Garrison}, {Gibbons}, {Goldstein}, {Gommers},
  {Greco}, {Greenfield}, {Groener}, {Grollier}, {Hagen}, {Hirst}, {Homeier},
  {Horton}, {Hosseinzadeh}, {Hu}, {Hunkeler}, {Ivezi{\'c}}, {Jain}, {Jenness},
  {Kanarek}, {Kendrew}, {Kern}, {Kerzendorf}, {Khvalko}, {King}, {Kirkby},
  {Kulkarni}, {Kumar}, {Lee}, {Lenz}, {Littlefair}, {Ma}, {Macleod},
  {Mastropietro}, {McCully}, {Montagnac}, {Morris}, {Mueller}, {Mumford},
  {Muna}, {Murphy}, {Nelson}, {Nguyen}, {Ninan}, {N{\"o}the}, {Ogaz}, {Oh},
  {Parejko}, {Parley}, {Pascual}, {Patil}, {Patil}, {Plunkett}, {Prochaska},
  {Rastogi}, {Reddy Janga}, {Sabater}, {Sakurikar}, {Seifert}, {Sherbert},
  {Sherwood-Taylor}, {Shih}, {Sick}, {Silbiger}, {Singanamalla}, {Singer},
  {Sladen}, {Sooley}, {Sornarajah}, {Streicher}, {Teuben}, {Thomas},
  {Tremblay}, {Turner}, {Terr{\'o}n}, {van Kerkwijk}, {de la Vega}, {Watkins},
  {Weaver}, {Whitmore}, {Woillez}, {Zabalza}, \& {Contributors}}]{astropy:2018}
{Price-Whelan}, A.~M., {Sip{\H{o}}cz}, B.~M., {G{\"u}nther}, H.~M., {et~al.}
  2018, \aj, 156, 123

\bibitem[{{R Core Team}(2015)}]{rstat}
{R Core Team}. 2015, R: A Language and Environment for Statistical Computing, R
  Foundation for Statistical Computing, Vienna, Austria

\bibitem[{{S{\'a}nchez Almeida} {et~al.}(2015){S{\'a}nchez Almeida},
  {Elmegreen}, {Mu{\~n}oz-Tu{\~n}{\'o}n}, {Elmegreen}, {P{\'e}rez-Montero},
  {Amor{\'{\i}}n}, {Filho}, {Ascasibar}, {Papaderos}, \&
  {V{\'{\i}}lchez}}]{SanchezAlmeida15}
{S{\'a}nchez Almeida}, J., {Elmegreen}, B.~G., {Mu{\~n}oz-Tu{\~n}{\'o}n}, C.,
  {et~al.} 2015, \apjl, 810, L15

\bibitem[{{S{\'a}nchez Almeida} {et~al.}(2013){S{\'a}nchez Almeida},
  {Mu{\~n}oz-Tu{\~n}{\'o}n}, {Elmegreen}, {Elmegreen}, \&
  {M{\'e}ndez-Abreu}}]{SanchezAlmeida13}
{S{\'a}nchez Almeida}, J., {Mu{\~n}oz-Tu{\~n}{\'o}n}, C., {Elmegreen}, D.~M.,
  {Elmegreen}, B.~G., \& {M{\'e}ndez-Abreu}, J. 2013, \apj, 767, 74

\bibitem[{{S{\'a}nchez-Janssen} {et~al.}(2010){S{\'a}nchez-Janssen},
  {M{\'e}ndez-Abreu}, \& {Aguerri}}]{Sanchez-Janssen10}
{S{\'a}nchez-Janssen}, R., {M{\'e}ndez-Abreu}, J., \& {Aguerri}, J.~A.~L. 2010,
  \mnras, 406, L65

\bibitem[{{Sersic}(1968)}]{Sersicdg}
{Sersic}, J.~L. 1968

\bibitem[{{Shibuya} {et~al.}(2015){Shibuya}, {Ouchi}, \&
  {Harikane}}]{Shibuya15}
{Shibuya}, T., {Ouchi}, M., \& {Harikane}, Y. 2015, \apjs, 219, 15

\bibitem[{{Skelton} {et~al.}(2014){Skelton}, {Whitaker}, {Momcheva}, {Brammer},
  {van Dokkum}, {Labb{\'e}}, {Franx}, {van der Wel}, {Bezanson}, {Da Cunha},
  {Fumagalli}, {F{\"o}rster Schreiber}, {Kriek}, {Leja}, {Lundgren}, {Magee},
  {Marchesini}, {Maseda}, {Nelson}, {Oesch}, {Pacifici}, {Patel}, {Price},
  {Rix}, {Tal}, {Wake}, \& {Wuyts}}]{Skelton14}
{Skelton}, R.~E., {Whitaker}, K.~E., {Momcheva}, I.~G., {et~al.} 2014, \apjs,
  214, 24

\bibitem[{{Smethurst} {et~al.}(2015){Smethurst}, {Lintott}, {Simmons},
  {Schawinski}, {Marshall}, {Bamford}, {Fortson}, {Kaviraj}, {Masters},
  {Melvin}, {Nichol}, {Skibba}, \& {Willett}}]{Smethurst15}
{Smethurst}, R.~J., {Lintott}, C.~J., {Simmons}, B.~D., {et~al.} 2015, \mnras,
  450, 435

\bibitem[{{Sparre} {et~al.}(2017){Sparre}, {Hayward}, {Feldmann},
  {Faucher-Gigu{\`e}re}, {Muratov}, {Kere{\v s}}, \& {Hopkins}}]{Sparre17}
{Sparre}, M., {Hayward}, C.~C., {Feldmann}, R., {et~al.} 2017, \mnras, 466, 88

\bibitem[{{Starkenburg} {et~al.}(2016){Starkenburg}, {Helmi}, \&
  {Sales}}]{Starkenburg16b}
{Starkenburg}, T.~K., {Helmi}, A., \& {Sales}, L.~V. 2016, \aap, 595, A56

\bibitem[{{Stierwalt} {et~al.}(2015){Stierwalt}, {Besla}, {Patton}, {Johnson},
  {Kallivayalil}, {Putman}, {Privon}, \& {Ross}}]{Stierwalt15}
{Stierwalt}, S., {Besla}, G., {Patton}, D., {et~al.} 2015, \apj, 805, 2

\bibitem[{{Taniguchi} {et~al.}(2015){Taniguchi}, {Kajisawa}, {Kobayashi},
  {Shioya}, {Nagao}, {Capak}, {Aussel}, {Ichikawa}, {Murayama}, {Scoville},
  {Ilbert}, {Salvato}, {Sanders}, {Mobasher}, {Miyazaki}, {Komiyama}, {Le
  F{\`e}vre}, {Tasca}, {Lilly}, {Carollo}, {Renzini}, {Rich}, {Schinnerer},
  {Kaifu}, {Karoji}, {Arimoto}, {Okamura}, {Ohta}, {Shimasaku}, \&
  {Hayashino}}]{Taniguchi15}
{Taniguchi}, Y., {Kajisawa}, M., {Kobayashi}, M.~A.~R., {et~al.} 2015, \pasj,
  67, 104

\bibitem[{{Tasca} {et~al.}(2015){Tasca}, {Le F{\`e}vre}, {Hathi}, {Schaerer},
  {Ilbert}, {Zamorani}, {Lemaux}, {Cassata}, {Garilli}, {Le Brun}, {Maccagni},
  {Pentericci}, {Thomas}, {Vanzella}, {Zucca}, {Amorin}, {Bardelli},
  {Cassar{\`a}}, {Castellano}, {Cimatti}, {Cucciati}, {Durkalec}, {Fontana},
  {Giavalisco}, {Grazian}, {Paltani}, {Ribeiro}, {Scodeggio}, {Sommariva},
  {Talia}, {Tresse}, {Vergani}, {Capak}, {Charlot}, {Contini}, {de la Torre},
  {Dunlop}, {Fotopoulou}, {Koekemoer}, {L{\'o}pez-Sanjuan}, {Mellier}, {Pforr},
  {Salvato}, {Scoville}, {Taniguchi}, \& {Wang}}]{Tasca15}
{Tasca}, L.~A.~M., {Le F{\`e}vre}, O., {Hathi}, N.~P., {et~al.} 2015, \aap,
  581, A54

\bibitem[{{Tolstoy} {et~al.}(2009){Tolstoy}, {Hill}, \& {Tosi}}]{Tolstoy09}
{Tolstoy}, E., {Hill}, V., \& {Tosi}, M. 2009, \araa, 47, 371

\bibitem[{{Tortora} {et~al.}(2010){Tortora}, {Napolitano}, {Cardone},
  {Capaccioli}, {Jetzer}, \& {Molinaro}}]{Tortora10}
{Tortora}, C., {Napolitano}, N.~R., {Cardone}, V.~F., {et~al.} 2010, \mnras,
  407, 144

\bibitem[{{van den Bergh} {et~al.}(1996){van den Bergh}, {Abraham}, {Ellis},
  {Tanvir}, {Santiago}, \& {Glazebrook}}]{vandenBergh96}
{van den Bergh}, S., {Abraham}, R.~G., {Ellis}, R.~S., {et~al.} 1996, \aj, 112,
  359

\bibitem[{{van der Wel} {et~al.}(2014){van der Wel}, {Franx}, {van Dokkum},
  {Skelton}, {Momcheva}, {Whitaker}, {Brammer}, {Bell}, {Rix}, {Wuyts},
  {Ferguson}, {Holden}, {Barro}, {Koekemoer}, {Chang}, {McGrath},
  {H{\"a}ussler}, {Dekel}, {Behroozi}, {Fumagalli}, {Leja}, {Lundgren},
  {Maseda}, {Nelson}, {Wake}, {Patel}, {Labb{\'e}}, {Faber}, {Grogin}, \&
  {Kocevski}}]{vanderwel14}
{van der Wel}, A., {Franx}, M., {van Dokkum}, P.~G., {et~al.} 2014, \apj, 788,
  28

\bibitem[{{Vulcani} {et~al.}(2014){Vulcani}, {Bamford}, {H{\"a}u{\ss}ler},
  {Vika}, {Rojas}, {Agius}, {Baldry}, {Bauer}, {Brown}, {Driver}, {Graham},
  {Kelvin}, {Liske}, {Loveday}, {Popescu}, {Robotham}, \& {Tuffs}}]{Vulcani14}
{Vulcani}, B., {Bamford}, S.~P., {H{\"a}u{\ss}ler}, B., {et~al.} 2014, \mnras,
  441, 1340

\bibitem[{{Wang} {et~al.}(2017){Wang}, {Kong}, {Wang}, {Wang}, {Lin}, {Gao}, \&
  {Liu}}]{Wang17}
{Wang}, E., {Kong}, X., {Wang}, H., {et~al.} 2017, \apj, 844, 144

\bibitem[{{Wang} {et~al.}(2018){Wang}, {Li}, {Xiao}, {Lin}, {Bershady}, {Law},
  {Merrifield}, {Sanchez}, {Riffel}, {Riffel}, \& {Yan}}]{Wang18}
{Wang}, E., {Li}, C., {Xiao}, T., {et~al.} 2018, \apj, 856, 137

\bibitem[{{Whitaker} {et~al.}(2014){Whitaker}, {Franx}, {Leja}, {van Dokkum},
  {Henry}, {Skelton}, {Fumagalli}, {Momcheva}, {Brammer}, {Labb{\'e}},
  {Nelson}, \& {Rigby}}]{Whitaker14}
{Whitaker}, K.~E., {Franx}, M., {Leja}, J., {et~al.} 2014, \apj, 795, 104

\bibitem[{{Wright} {et~al.}(2019){Wright}, {Brooks}, {Weisz}, \&
  {Christensen}}]{Wright19}
{Wright}, A.~C., {Brooks}, A.~M., {Weisz}, D.~R., \& {Christensen}, C.~R. 2019,
  \mnras, 482, 1176

\bibitem[{{Wuyts} {et~al.}(2011){Wuyts}, {F{\"o}rster Schreiber}, {van der
  Wel}, {Magnelli}, {Guo}, {Genzel}, {Lutz}, {Aussel}, {Barro}, {Berta},
  {Cava}, {Graci{\'a}-Carpio}, {Hathi}, {Huang}, {Kocevski}, {Koekemoer},
  {Lee}, {Le Floc'h}, {McGrath}, {Nordon}, {Popesso}, {Pozzi}, {Riguccini},
  {Rodighiero}, {Saintonge}, \& {Tacconi}}]{Wuyts11b}
{Wuyts}, S., {F{\"o}rster Schreiber}, N.~M., {van der Wel}, A., {et~al.} 2011,
  \apj, 742, 96

\end{thebibliography}

\begin{appendix}
\section{Creating the H$\alpha$ images}
\label{appen_imag}

In this appendix we explain in detail the process we followed to create the H$\alpha$ images, depending on the redshift of the galaxy.

\subsection{z<0.266 and z>0.298}
For galaxies with $z$<0.266 or $z$>0.298, the process is quite straightforward. Band A (the one with H$\alpha$ contamination) is F775W in the lower redshift case and F850LP in the higher redshift one. Band B is the one with only stellar continuum in its wavelength range (F850LP in the lower redshift case and F775W in the higher redshift one). We have then: 
\begin{equation}
    F_{A}=C_A+H\alpha
\end{equation}

\begin{equation}
  F_{B}=C_B  
\end{equation} 
where $F_A$ and $F_B$ are the total fluxes in each band, $C_A$ and $C_B$ the continua, and H$\alpha$ the line flux. Subtracting, and taking into account the scaling factor $f=C_A/C_B$ between the continua computed from the SHARDS SED fits in Paper I we get:
\begin{equation}
F_{A}-f\times F_{B}=H\alpha + C_A-f \times C_B=H\alpha
\end{equation}

\subsection{0.266<z<0.298}

For galaxies with 0.266<$z$<0.298, both F775W and F850LP are contaminated by H$\alpha$, therefore the previous approach is unfeasible. The use of the F606W filter as continuum is impossible because it is affected by [OIII] and H$\beta$ contamination (and other filters would be too distant in wavelength). We consider then:
\begin{equation}
    F_{F775W}=C_{F775W}+H\alpha
\end{equation}
\begin{equation}
    F_{F850LP}=C_{F850LP}+H\alpha
\end{equation}

We have then:
\begin{equation}
    F_{F775W}-F_{F850LP}=C_{F775W}-C_{F850LP}=S
\end{equation}
Computing the scaling factor $f=C_{F775W}/C_{F850LP}$ between the continua:
\begin{equation}
    f\times C_{F850LP} - C_{F850LP} = Ima_3 \rightarrow C_{F850LP}=\frac{S}{f-1}
\end{equation}

Substituting in the second equation, we get:
\begin{equation}
    H\alpha=F_{850LP}-\frac{S}{f-1}
\end{equation}

\end{appendix}
\end{document}